%% file: physicaD.tex
\documentclass[11pt]{article}
\usepackage{epsfig}
\psfigdriver{dvips}
\newtheorem{theo}{Theorem}[section]

\newtheorem{prop}[theo]{Proposition}

\newtheorem{conj}[theo]{Conjecture}
\newtheorem{defi}[theo]{Definition}

\input{epsf}

\begin{document}

\def\C{{\rm\kern.24em \vrule width.02em height1.4ex
depth-.05ex\kern-.26em C}}
\def\N{{\rm I\hspace{-0.4ex}N}}
\def\R{{\rm I\hspace{-0.4ex}R}}
\def\rit{{\rm I\hspace{-0.4ex}R}}
\def\grad{\hbox{ grad }}
\def\curl{\hbox{curl }}
\def\div{\hbox{ div }}
\def\supp{{\rm supp}\:}
\def\Lip{{\rm Lip}}
\def\sign{{\rm sign}\:}
\def\dist{{\rm dist}}
\def\diam{{\rm diam}\:}
\def\const{{\rm const.}\:}
\def\meas{{\rm meas}\:}
\def\oOmega{\overline{\Omega}}
\def\eps{\varepsilon}
\def\be{\begin{equation}}
\def\ee{\end{equation}}
\def\beq{\begin{equation}}
\def\eeq{\end{equation}}
\def\ds{\displaystyle}
\def\ts{\textstyle}
\def\qed{\rule{1ex}{1ex}}

\title{\bf  On the solutions of the one dimensional Ginzburg-Landau equations for superconductivity}
\author{Amandine Aftalion\footnote{DMI, Ecole Normale Sup\'erieure, 45, rue d'Ulm, 75230 Paris cedex 05, France. }$\ $
  and William C. Troy\footnote{Mathematics departement, University of Pittsburgh, Pittsburgh, Pennsylvania 15 260, USA.}}
\maketitle

\begin{abstract}
This paper gives a complete description of the solutions of the
 one dimensional Ginzburg-Landau equations which model
 superconductivity phenomena in infinite slabs. We investigate this problem over the entire range of physically important parameters: $a$ the size of the slab, $\kappa$ the Ginzburg-Landau parameter, and $h_0$, the exterior magnetic field.
 We do extensive numerical computations using the software AUTO, and
 determine the number, symmetry and stability of solutions for all values of the parameters.
 In particular, our experiments reveal the existence of two key-points in parameter space which play a central role
 in the formation of the complicated patterns by means of bifurcation phenomena. Our global
 description also allows us to separate the various physically important
 regimes, to classify previous results in each regime according to the values of the parameters and to derive new open problems.
 In addition, our investigation provides new insight into 
 the problem of differentiating
 between the types of superconductors in terms of the parameters.
\end{abstract}

\noindent 
{\bf PACS NUMBERS:} 2.60.Lj, 5.70.Fh, 74.20.De

\smallskip


\section{Introduction}
\setcounter{equation}{0}

This paper gives a complete description of the solutions of the
 one dimensional Ginzburg-Landau equations which model
 superconductivity phenomena in infinite slabs. One of our main goals here is
 to determine the number, symmetry and stability of solutions. We give a complete numerical investigation of these issues over the entire range of physically important parameters:
$a$ the size of the slab, $\kappa$ the Ginzburg-Landau parameter, and $h_0$, the exterior magnetic field.
 Our experiments, which are summarized in Figure \ref{figsummary}, reveal the existence of two key-points in parameter space which play a central role
 in the way superconductivity is nucleated. Indeed, according to the values of $a$ and $\kappa$, superconductivity appears either in the volume of the sample if the slab is thin, or on superconducting sheaths near the boundary if the slab is large, and in the form of two-dimensionnal vortex patterns if the slab is of intermediate size. This is why all values of $a$ and $\kappa$ are of interest.
 As we shall see, sheaths and vortices can be described in terms of asymmetric solutions of the one dimensional problem, 
 and each of the key-points gives rise to
 diverse pattern formation by means of  various bifurcation phenomena. 
   In addition, our results provide new insight into the problem of differentiating
 between the types of superconductors in terms of the parameters. Finally,
 our global
 description allows us to understand better many interesting physical phenomena, to classify results previously obtained for each regime of the parameters
 and also to derive new open problems.
  In order to properly
 describe our results we begin with a brief summary of the model.

\hfill

 The superconductivity of certain metals is characterized at very low temperatures
 by the loss of electrical resistance and the expulsion of the exterior magnetic
 field $\hbox{\bf h}_0$.
 In the model derived by Ginzburg and Landau in 1950
 (see \cite{GL}), the electromagnetic properties of the
 material are completely described by
 the magnetic potential vector $\hbox{\bf A}$
 ($ \hbox{\bf h}=\curl \hbox{\bf A}$ being the magnetic field) and
 the complex-valued order parameter $\psi$.
 In fact, $\psi$ is an averaged wave function of
 the superconducting electrons and its modulus corresponds
 to the density of superconducting
 carriers. When the sample is wholly normal,
 $|\psi|\equiv 0$ and the magnetic field inside the material $\hbox{\bf h}$
 is equal to the exterior magnetic field $\hbox{\bf h}_0$. On the other
 hand, when the sample is perfectly superconducting,
 $|\psi|\equiv 1$ and the magnetic field $\hbox{\bf h}$ is identically 0.
 Furthermore, in the Ginzburg-Landau theory,
 the state of the sample is completely determined by the minimum
 of an energy depending on $\psi$ and $\hbox{\bf A}$.
 For a more precise description of the general theory, one may refer to
  \cite{CP}, \cite{DG},  \cite{GL}, \cite{SJDG}, \cite{TT}, \cite{T} or to
 \cite{CHO}, \cite{du}.

 In the special case when the sample is an infinite slab of
 constant thickness,
 between the planes $x=-a$ and $x=a$, it is usual to assume that
 both $\psi$ and {\bf A} are uniform in the $y$ and $z$ directions, and that
 the exterior magnetic field is
 tangential to the slab, that is $\hbox{\bf h$_0$}=$(0,0,$h_0$).
 A suitable gauge can then be chosen so that $\psi=f(x)$ is a real function, and 
 $\hbox{\bf A}=q(x)\hbox{\bf e}_y$, where {\bf e}$_y$ is the unit vector
 along the $y$ direction (see \cite{GL} for more details).
 In this case, the nondimensionalized form of the Ginzburg-Landau energy is given by: 
\beq \label{Ekappa} 
E_{\kappa}(f,q)=\int_{-a}^{a} \Bigl({1 \over {\kappa^2}}f'^2+f^2q^2+
{1\over2}f^4-f^2+(q'- h_0)^2 \Bigr )\ dx.
\eeq
The nondimensionalized parameter $\kappa$ is called the
 Ginzburg-Landau parameter.
 It is the ratio of $\lambda$, the penetration depth of the magnetic field, to
 $\xi$, the coherence length, which is the characteristic length of variation of $f$.
 The value of $\kappa$ determines the type of superconductor according to the
 type of phase transition which takes place between the normal phase and the
 superconducting phase: $\kappa$
 small describes what is known as a type I superconductor and $\kappa$ large
 as a type II. More precisely, for a type I superconductor, there is a critical
 magnetic field $h_c$ such that if $h_0<h_c$, the material is entirely superconducting
 and the magnetic field is expelled from the sample apart from a boundary layer
 of size $\lambda$. This is called the Meissner effect. If $h_0>h_c$, superconductivity
 is destroyed and the material is in the normal state, that is $f\equiv 0$
 and $q'\equiv h_0$. For a type II superconductor, the phase transition is different
 and there are two critical fields $h_{c_1}$ and $h_{c_2}$:
 for $h_0<h_{c_1}$, the exterior magnetic field is expelled from the sample and there is a Meissner effect as for type I superconductors.
 But as $h_0$ is increased above $h_{c_1}$, superconductivity is not destroyed  straight away, since the superconducting and the normal
 phase coexist under the form of filaments or vortices: the vortex is a zone
 of diameter $\xi$, at the center of which the order parameter $f$ vanishes. 
 As $h_0$ increases further, the vortices become more numerous until the critical value $h_{c_2}$ is reached at which superconductivity is destroyed.
 For $h_0>h_{c_2}$, there is no superconductivity
 and the material is in the normal state. The way superconductivity is nucleated is highly dependent on $a$ and $\kappa$, as we will see later in Section 4, where we will introduce a third critical field $h_{c_3}$ corresponding to nucleation of  surface superconductivity. We refer to Tinkham \cite{T} for a detailed explanation.
 The vortex phenomena in superconductivity have been widely studied in the literature. See for instance
 \cite{BBH} or \cite{du} and the references therein.
 The critical value of $\kappa$ usually given to separate type I
 and type II superconductors is $\kappa=1/\sqrt 2$. In the next Sections, we will
 describe how this value is computed in the limiting case $a=\infty$. However,
 as $a$ is decreased from infinity, we will see that the demarcation between
 type I and II behaviours is no longer the constant $\kappa= 1/ \sqrt 2$.
 Instead, we find that there is a well defined curve in the $(\kappa,a)$ plane
 which separates the two types of behaviours. This result will be fully discussed in Section 3. Similarly in Section 4, we find an extra curve which separates different
 type I superconductors according to surface effects.

\hfill

For a mathematical analysis of the problem, it is natural to assume that
 $f \in H^1(-a,a)$ and $q \in H^1(-a,a)$. It then follows from standard variational arguments
 that there exists a minimizer of $E_\kappa$, and that the minimizer is a solution of
$$\left\{\begin{array}{ll}
{1 \over \kappa^2}f''=f(f^2+q^2-1)
\quad\hbox{in}\quad (-a,a),\\
f'(\pm a)=0,\\
q''=qf^2 \quad\hbox{in}\quad (-a,a),\\
q'(\pm a)=h_0.
\end{array}\right.\eqno{(GL)}$$
Notice that $f\equiv 0$ and $q(x)=h_0(x+e)$ is always a solution for
 any real $e$. From now on, we will call this a {\em normal solution}.
 Regularity properties of minimizers yield that either $f$ is a normal
 solution, or $f$ does not change sign, hence we will study the case $f>0$. An easy calculation
 shows that the energy $E_\kappa$ is zero along the normal solution. Thus a global minimizer
 cannot have positive energy. 

\hfill 

The aim of this paper is to give a complete description (number, symmetry and stability) of the solutions
 of the system $(GL)$ for which $f>0$ on $[-a,a]$, according to the 
values of the parameters $a$, $\kappa$ and $h_0$. This work is based on numerical simulations obtained with AUTO, a software developed by Doedel et al. (see
 \cite{Doedel1}, \cite{Doedel2}, \cite{Doedel3}) which computes bifurcation diagrams for systems of ODE's.

In order to properly describe our results, it is necessary that we first give
 a brief summary of previous studies of the $(GL)$ system. This is done in
 the next section. Then, in Sections 3 and 4 we give the details of our
 numerical investigation for symmetric solutions (Section 3) and asymmetric solutions (Section 4), together with their physical implications and
 suggestions for future research. Our results are all summarized in Figure \ref{figsummary}.
 Section 5 is devoted to the stability analysis.

\section{Mathematical background}

Let us first recall the basic properties of solutions.
\begin{prop}\label{summary}
If $(f,q)$ is a solution of $(GL)$ and
 if f is not identically zero then
\begin {itemize}
\item[(i)]
$|f|\leq 1$ in $(-a,a)$,
\item[(ii)]
$q$ has a unique zero $a_0$ in $(-a,a)$,
 $q$ is increasing on $(-a,a)$,
 $q'$ is decreasing on $(-a,a_0)$ and increasing on $(a_0,a)$.
\item[(iii)] There exist $x_1$ and $x_2$ with $-a\leq
x_1\leq a_0\leq x_2\leq a$ and $x_0\in[x_1,x_2]$ such that
$f'$ is increasing on $[-a,x_1] \cup [x_2,a]$ and decreasing on $[x_1,x_2]$, 
 $f$ is increasing on $[-a,x_0]$ and decreasing on $[x_0,a]$.
\end{itemize}
\end{prop}
The proof of $(i)$ and $(ii)$ can be found for instance in \cite{BH3} and of $(iii)$
 in \cite{aa1}.

\hfill

There are two types of physically important solutions of $(GL)$: symmetric solutions and asymmetric solutions. We  define a {\em symmetric solution} to be
 a solution of $(GL)$ such that $f>0$, $f$
 is even and $q$ is odd on $[-a,a]$. Thus, a symmetric solution satisfies the following problem:
$$\left\{\begin{array}{ll}
{1 \over \kappa^2}f''=f(f^2+q^2-1)
\quad\hbox{in}\quad (0,a),\\
f(0)=\beta,\quad f'(0)=0,\\
q''=qf^2 \quad\hbox{in}\quad (0,a),\\
q(0)=0,\quad q'(0)=\alpha,
\end{array}\right.\eqno{(GL_{sym})}$$
for $\beta\in(0,1)$ and $\alpha\geq 0$. We need to choose $\alpha$ and
 $\beta$ such that $f'(a)=0$. Then $(f,q)$ will be a solution of $(GL)$ with
 $h_0=q'(a)$. Notice that $\beta$ is the amplitude of $f$ for a
  symmetric solution. 

We define an {\em asymmetric solution} to be a solution of
 $(GL)$ which satisfies $f>0$ on $[-a,a]$, yet
 which is not symmetric. That is $f'(0)\neq 0$ or $q(0)\neq 0$.

For the existence of symmetric solutions, Kwong \cite{kwong} has proved the  
 following important result.
\begin{theo}\label{kwong}
(Kwong \cite{kwong}) For each $\beta$ in $(0,1)$, there exists a unique
 $\alpha>0$ such that the solution $(f,q)$ of $(GL_{sym})$ satisfies $f'(a)=0$.
 Moreover, $\alpha$ is a continuous, decreasing function of $\beta$,
\beq\label{kwongalpha}
\lim_{\beta\to 0}\alpha(\beta)>0\quad \hbox{and}\quad \lim_{\beta\to 1}\alpha(\beta)=0.
\eeq 
For this choice of $\alpha(\beta)$, and the corresponding
 solution $(f,q)$ of $(GL_{sym})$, let $h(\beta)=q'(a)$. Then $h$ is well-defined, continuous and
 \beq\label{kwongh}
\lim_{\beta\to 0}h(\beta)=h_s>0\quad \hbox{and}\quad \lim_{\beta\to 1}h(\beta)=0.
\eeq 
\end{theo}
In the following the notation $h_s$ of (\ref{kwongh}) plays an important role
 in the discussion of our results.

\begin{figure}[p]
\vspace{-1cm}
$$\epsfxsize=9cm
\epsfysize=5.5cm
\epsfbox{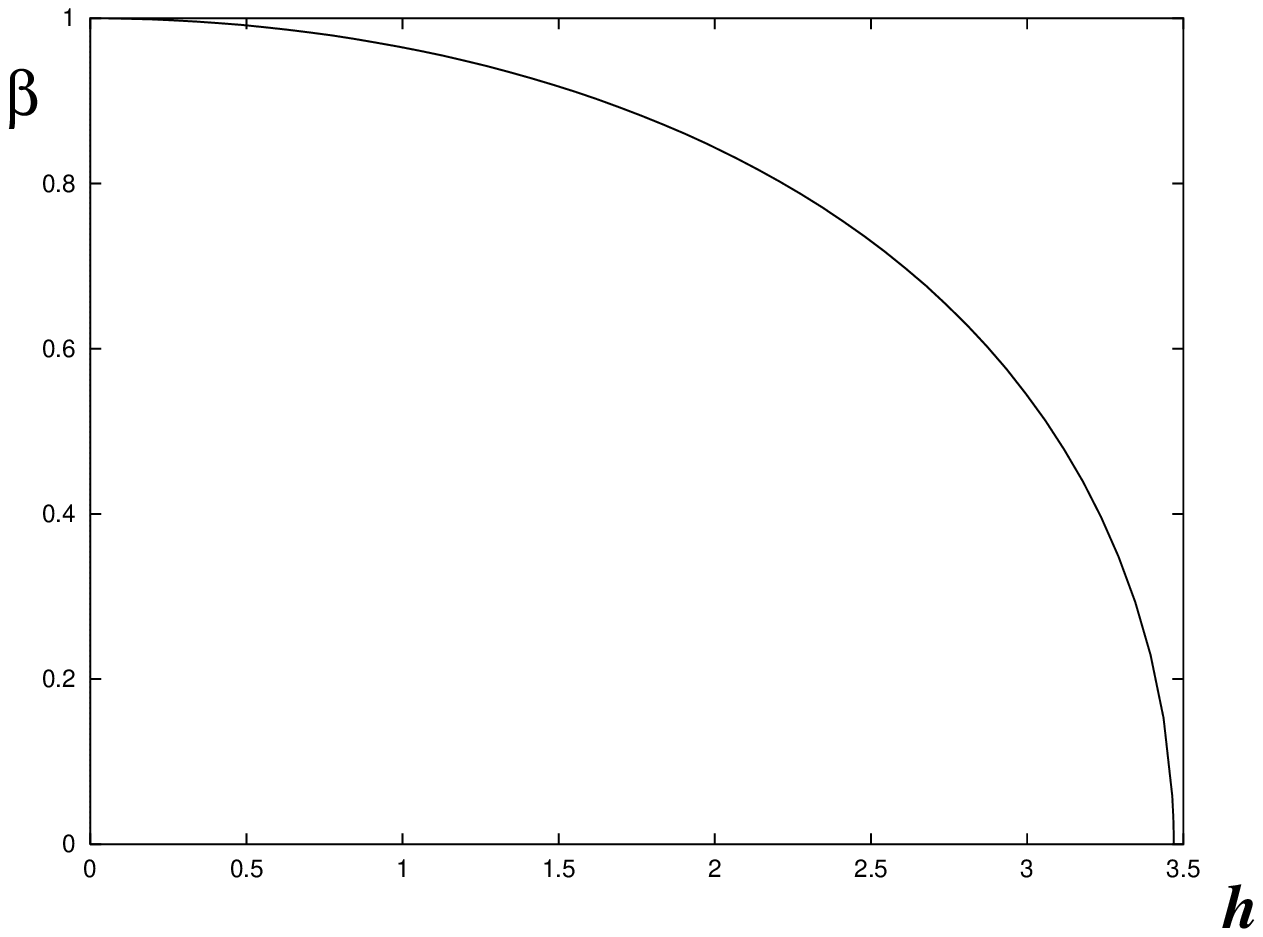}$$
\vspace{-1.2cm}
\caption{Curve $h(\beta)$ for $a=0.5$ and $\kappa=0.4$}
\label{figS1}
$$\epsfxsize=9cm
\epsfysize=5.5cm
\epsfbox{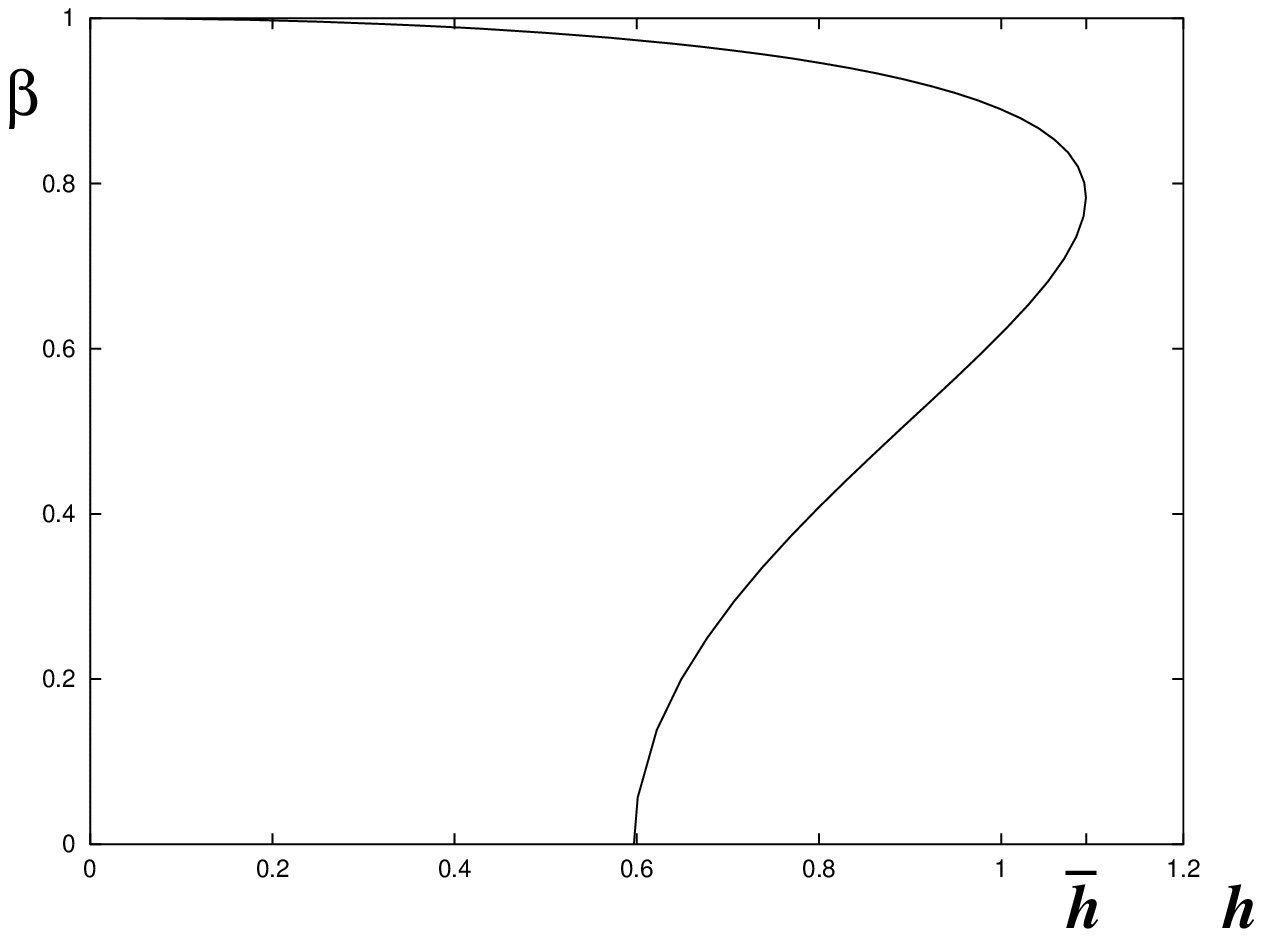}$$
\vspace{-1.2cm}
\caption{Curve $h(\beta)$ for $a=3$ and $\kappa=0.3$}
\label{figS2}
$$\epsfxsize=9cm
\epsfysize=5.5cm
\epsfbox{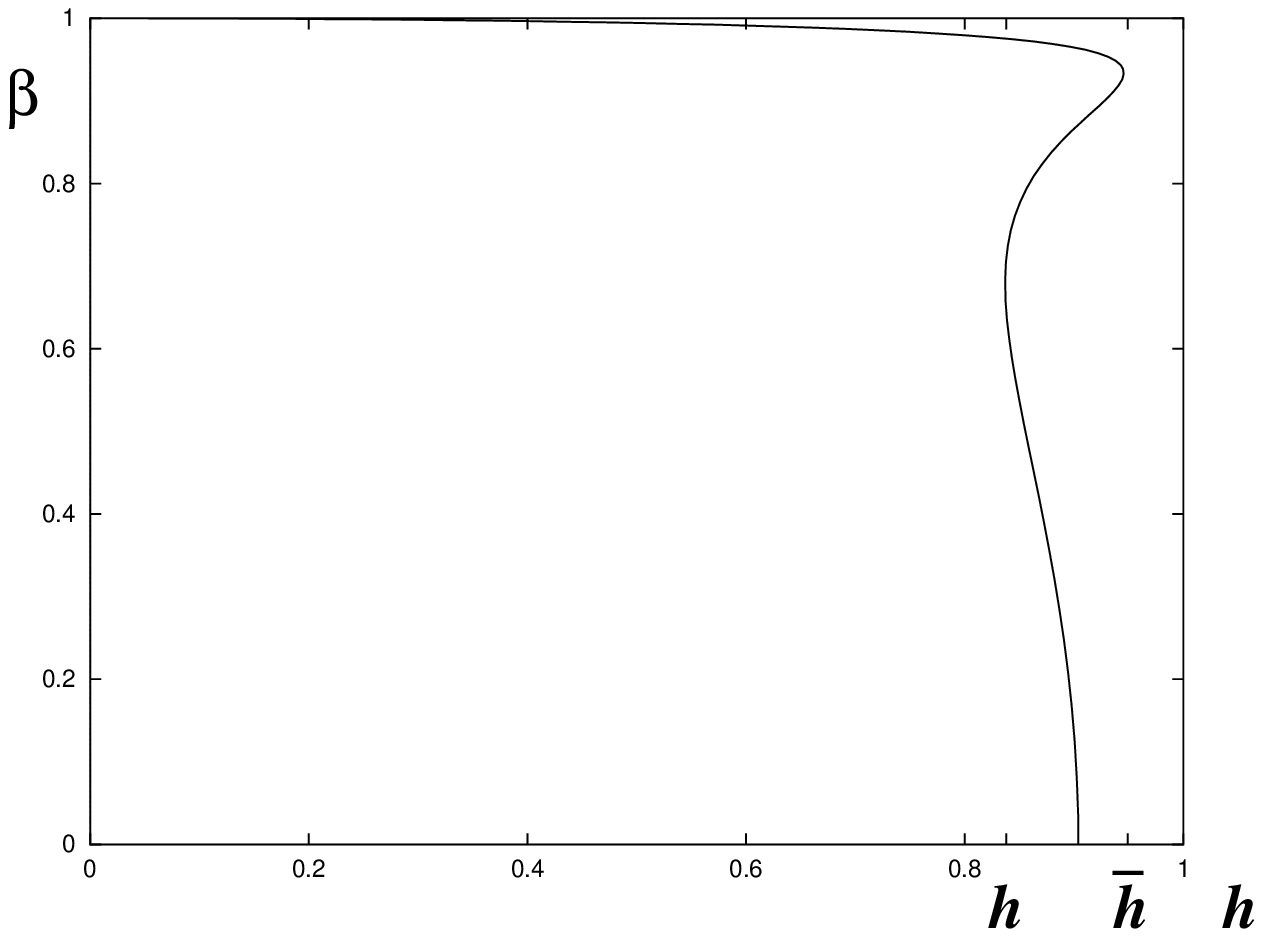}$$
\vspace{-1.2cm}
\caption{Curve $h(\beta)$ for $a=3$ and $\kappa=0.9$}
\label{figS3}
\end{figure}

 There are three possible behaviours of the curve $h(\beta)$ defined in Kwong's Theorem, and these are shown
 in Figures \ref{figS1}, \ref{figS2} and \ref{figS3}.
 Notice that
 instead of graphing $h(\beta)$ vs $\beta$, we have put $\beta$ on the
 vertical axis and $h$ on the horizontal axis, so that we keep the convention
 originally adopted by Ginzburg \cite[Figures 3 and 5]{G}.
\begin{itemize}
\item Figure \ref{figS1}: for $\beta \in (0,1)$, $h(\beta)$
 is a decreasing function of $\beta$: if $0<h_0<h_s$, there is exactly
 one symmetric solution of $(GL)$, and if $h_0\geq h_s$, no such solution.
\item Figure \ref{figS2}: for $\beta \in (0,1)$, $h(\beta)$ is increasing until $h$ reaches
 a maximum value $\overline{h}$ (here $\overline{h}\simeq 1.1$),
 and then decreasing to 0 as $\beta$ goes to 1.
 Thus, for $h_0\leq h_{s}$, $(GL)$ has a unique symmetric solution,
 for $h_s< h_0<\overline{h}$, two symmetric solutions,
 and
 for $h_0> \overline{h}$ no such solution. 

\item Figure \ref{figS3}: for $\beta \in (0,1)$, $h(\beta)$ is decreasing until $h$ reaches
 a local minimum value $\underline{h}$ (here $\underline{h}\simeq 0.83$), increasing until $h$ reaches a local maximum value
 $\overline{h}$ (here $\overline{h}\simeq 0.95$), 
 and then decreasing to 0 as $\beta$ goes to 1.
 There $\underline{h}<h_s< \overline{h}$. Thus,
 for $h_0<\underline{h}$, $(GL)$ has a unique symmetric solution,
 for $\underline{h}< h_0<{h}_{s}$, three
 symmetric solutions,
 for ${h}_s\leq h_0<\overline{h}$, two
 symmetric solutions,
 and
 for $h_0> \overline{h}$ no symmetric solution.

If $\overline{h}< {h}_{s}$, then
 for $h_0<\underline{h}$, $(GL)$ has a unique symmetric solution,
 for $\underline{h}< h_0<\overline{h}$, three
 symmetric solutions,
 for $\overline{h}< h_0<{h}_{s}$, a unique
 symmetric solution,
 and
 for $h_0\geq {h}_s$ no symmetric solution.
\end{itemize}
Note that in Figure \ref{figS2} and \ref{figS3},  there are points on the 
 curve where the resultant slope is vertical, when $h=\underline{h}$ or $\overline{h}$.
 We shall refer to such points as {\em folds} throughout the paper.

These figures indicate (and it has been verified experimentally)
 that if the applied field  $h_0$ is sufficiently 
 strong, superconductivity is destroyed.
 This has been proved by
 Kwong \cite{kwong}.
\begin{prop}\label{meissner}
(Kwong \cite{kwong})
For any $a$ and $\kappa$, there exists $h_c$ such that for $h_0>h_c$, the only
 solution of $(GL)$ is the normal solution $f\equiv 0$, $q'\equiv h_0$.
\end{prop}
A weaker result had been obtained earlier in \cite{wang} and \cite{Y}.

 As each of the Figures indicates, as one decreases $h_0$ from infinity, the material remains in the normal
 state until a critical value of $h_0$ is reached at which
 there is a
 bifurcation of nontrivial solutions from the normal state. 
 So for small $\eps>0$, a nontrivial curve $(f(.,\eps),q(.,\eps),h(\eps))$ of solutions of $(GL)$
 starting from a normal solution $(0,h_0(x+e),h_0)$ is sought, with
 the following asymptotic development:
\beq\label{bifurcatedsol}
\left.\begin{array}{ll}
f(x,\eps)=\eps f_0(x)+\eps^3 f_1(x) +o(\eps^3)\quad \hbox{in} \quad
H^2(-a,a),\\
q(x,\eps)=q_0(x)+\eps^2 q_1(x) +o(\eps^2)\quad \hbox{in} \quad
H^2(-a,a),\\
h(\eps)=h_0+\eps^2 h_1+\eps^4 h_2+o(\eps^4),
\end{array}\right.
\eeq 
 where $q_0(x)=h_0(x+e)$. It is important to note that when $e=0$, the branch gives rise to symmetric
 solutions and when $e\neq 0$ to asymmetric solutions. 

 Furthermore, it can be proved (see \cite{chapmanI} or \cite{BH2} for more details)
 that for small $\eps>0$, the energy of the bifurcated superconducting solution,
 $E_\kappa(f(.,\eps), q(.,\eps))$, has the same sign
 as $h_1$. This leads us to the following definition.
\begin{defi}\label{bifur}
A curve of the form (\ref{bifurcatedsol}) is said to result from a supercritical bifurcation 
 if $h_1<0$. Then, for small $\eps>0$, the bifurcated solutions have lower energy than the normal state.

A bifurcation is said to be 
 subcritical if $h_1>0$. Here the bifurcated solutions have
 larger energy than the normal state.
\end{defi} 
This means that for Figures \ref{figS1} and \ref{figS3}, the bifurcation
 is supercritical and for Figure \ref{figS2} subcritical. 

We now summarize
 the main results previously obtained concerning these figures describing symmetric solutions.
 We note that Ginzburg \cite{G} had investigated the case $\kappa$ small
 and found that $h(\beta)$ behaves as in Figure \ref{figS1} for small $a$
 and, as $a$ is increased through a critical value, the graph of $h(\beta)$
 changes from Figure \ref{figS1}
 to \ref{figS2}. He explained the type of behaviour described by Figure \ref{figS2} in terms of superheating and supercooling. More precisely,
 when $h_0$ is large, superconductivity does not occur and the material
 is in the normal state. As $h_0$ is decreased, the material stays in the normal
 state down to $h_s$, even though there is a range of $h_0$ where the normal
 solution is only a local minimizer and the global minimizer is a superconducting solution (see \cite[Figure 2]{G}). If $h_0$ is decreased further, there is a jump in the maximum of $f$ and the material becomes superconducting, the solution being given by the symmetric branch. In this case, $h_s$ is called the
 supercooling field. Now, on the contrary, start from $h_0=0$ where the
 superconducting state $(1,0)$ is the global minimizer and increase $h_0$.
 The material will remain superconducting until $\overline{h}$ is reached,
 though for fields slightly less then $\overline{h}$, it is only a local minimizer and the global minimizer is the normal solution. For $h_0$ above $\overline{h}$, there is a jump in the maximum of $f$ and the material reverts to the normal state. This is the superheating phenomenon.
 These two phenomena give rise to a hysteresis loop as described in \cite{G} or
 later in \cite{du}. We will make further remarks about these stability properties in Section 5. 

Ginzburg and Landau \cite{GL} had also noticed that if $a$ large is fixed,
 then there occurs a symmetric supercritical bifurcation of superconducting solutions
 from the normal state
 as $\kappa$ is increased through a critical value, but had no special explanation for this, since at that time only superconductors with small $\kappa$ were known. However, as we shall see in the next Section, as $\kappa$ is increased
 through this critical value, our studies indicate that the behaviour seen in
 Figure \ref{figS2} changes into that seen in Figure \ref{figS3}.

Chapman \cite{chapmanI} has studied the case $a=\infty$, and showed that there
 is a change of bifurcation from subcritical to supercritical that takes
 place for $\kappa=1/\sqrt 2$, which is the critical value between type I and II superconductors. 
 Moreover, in \cite{chapmanII}, a linear stability analysis through the time
 dependent equations yield that the value $\kappa=1/\sqrt 2$ is also
 the one for which stability of the normal solution switches.
 In \cite{chap2}, he has investigated the behaviour
 of the superheating field $\overline{h}$ in the limit $a=\infty$.

Following these works which are mainly based on formal computations, Bolley and Helffer have
  extensively studied the phenomenon of bifurcation of solutions from the normal
 solution in \cite{bol1}, \cite{BH1}, \cite{BH2}, \cite{BH3}, \cite{BH4}, \cite{BH7}, \cite{BH8}. In particular, they have given rigorous proofs of properties of bifurcating branches and asymptotic formula for the
 superheating and supercooling fields. Although their results are mainly local, they
 have made a first attempt in \cite{BH8} to give a global stability picture of the solutions.  We will mention the details
 of their contributions relevant to the values of the parameters in the course
 of the discussion, and indicate how their results contribute to the broader global picture.

 Recently, Hastings, Kwong and Troy \cite{HKT} have further investigated the multiplicity
 of symmetric solutions. They have proved that if $\kappa\leq 1/\sqrt 2$, and $a$
 is sufficiently large, then the behaviour of $h(\beta)$ is described by Figure \ref{figS2}, and if $\kappa> 1/\sqrt 2$ and $a$ sufficiently
 large, by Figure \ref{figS3}.

Seydel \cite{S1} \cite{S2} was the first to give numerical evidence that there is
 a range of parameters for which asymmetric solutions and multiple
 symmetric solutions coexist. 


\hfill




One of the major goals of this paper is to extend the results described above
 and completely determine the multiplicity and stability of symmetric and asymmetric
 solutions for the
 entire range of values of $\kappa$ and $a$. In the next two sections, we present the detail of our numerical investigations for symmetric solutions (Section 3) and asymmetric solutions (Section 4), together with their physical implications and suggestions for future research. In particular in Section 4, we explain  how asymmetric solutions of $(GL)$ give rise to nucleation of sheaths and vortices. The last Section is devoted to stability properties.

\section{Symmetric solutions}
\setcounter{equation}{0}

\begin{figure}[htb]
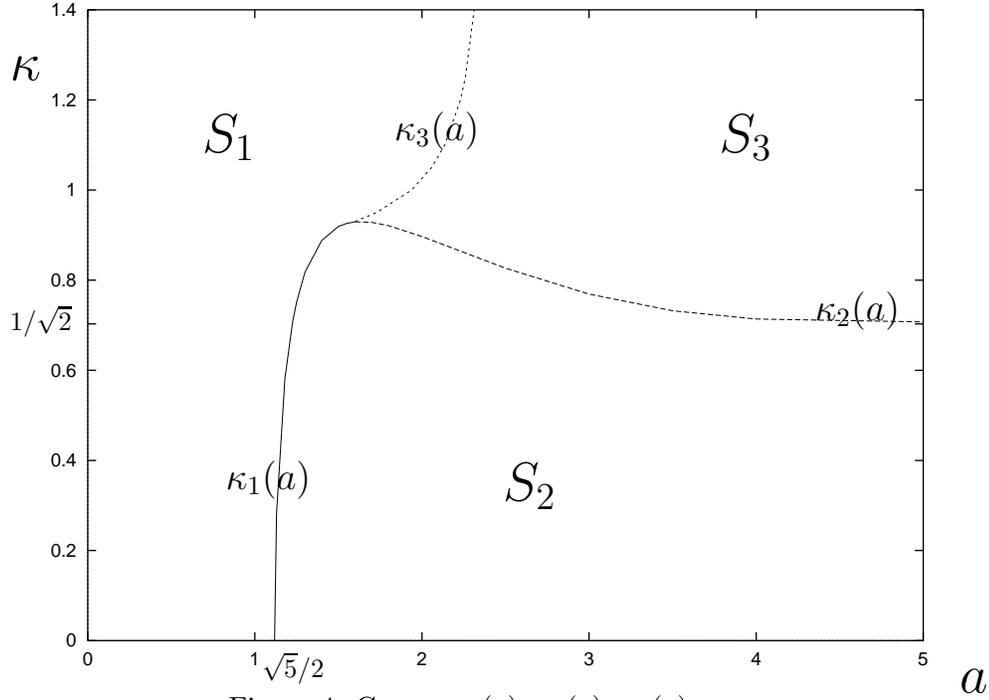

\begin{center}
 \input autoka.pstex_t 
\end{center}
\vspace{-1cm}
\caption{Curves $\kappa_1(a)$, $\kappa_2(a)$, $\kappa_3(a)$.}
\label{figkappaasym}
\end{figure}

As we stated in the last section, one of the main goals of this paper is to gain a complete understanding
 of the important properties of symmetric solutions in terms of the parameters
 $\kappa$ and $a$. 
 As we shall see, these global properties give us new insight into an important
 problem, that of demarcating the $(a,\kappa)$ plane into regions corresponding
 to type I and type II behaviours. The results of our numerical investigations concerning the existence and multiplicity of symmetric solutions are shown graphically in
 Figure \ref{figkappaasym}. 
 They indicate that the $(a,\kappa)$ plane is the 
 union of three connected sets $S_1$, $S_2$ and $S_3$. In $S_1$, the behaviour
 of $h(\beta)$ of Figure \ref{figS1} holds. Likewise, $S_2$ and $S_3$ reflect
 the behaviour
 of $h(\beta)$ of Figures \ref{figS2} and \ref{figS3} respectively.
 In the following, we explain the results shown in Figure \ref{figkappaasym} and make
 several conjectures.
\begin{conj}\label{main}
There exist three continuous functions  $\kappa_1(a)$, $\kappa_2(a)$
 and $\kappa_3(a)$ which separate the $(a,\kappa)$ plane into three
 connected regions
 $S_1$, $S_2$ and $S_3$, as shown in Figure \ref{figkappaasym}. 
 There exists exactly one point
 $a_*$  called the triple point (with approximate value 1.60) such that $\kappa_1(a_*)=\kappa_2(a_*)
=\kappa_3(a_*)=\kappa_* \simeq 0.93$. Moreover, $\kappa_1(a)$ is defined and monotone
 increasing on $[\sqrt5 /2, a_*]$, $\kappa_3(a)$ is defined and monotone increasing on
  $[a_*,\infty)$ and  $\kappa_2(a)$ is defined and monotone decreasing on
 $[a_*,\infty)$, 
\beq\label{kappa(a)}
\lim_{a\to \sqrt 5 /2}\kappa_1(a)=0, \quad
\lim_{a\to \infty}\kappa_2(a)=1/\sqrt 2,\quad
\lim_{a\to \infty}\kappa_3(a)=+\infty.
\eeq
\begin{itemize}
\item region $S_1=\{0<a\leq \sqrt 5/2,\kappa>0\}\cup\{\sqrt 5/2<a
 \leq a_*,\kappa\geq\kappa_1(a)\}\cup \{a>a_*,\kappa\geq\kappa_3(a)\}$.
 If $(a,\kappa)$ is in $S_1$, the behaviour of the corresponding curve $h(\beta)$ is given
 by Figure \ref{figS1} and
 there is at most one symmetric solution of $(GL)$. 
\item region $S_2=\{\sqrt 5/2<a < a_*,0<\kappa<\kappa_1(a)\}\cup \{a>a_*,
0<\kappa\leq\kappa_2(a)\}$. If $(a,\kappa)$ is in $S_2$,
 the behaviour of the corresponding curve $h(\beta)$ is given
 by Figure \ref{figS2} and 
 there are at most two symmetric solutions of $(GL)$. 

Let us call $\overline{\beta}$ the point where $h(\overline{\beta})=
\overline{h}$. Then as $(a,\kappa)$ tends to a point 
 $(\tilde{a},\kappa_1(\tilde{a}))$ on the curve $\kappa_1(a)$, $(\overline{\beta},\overline{h})$
 tends to $(0,h_s)$.

\item region $S_3=\{a> a_*,\kappa_2(a)<\kappa<\kappa_3(a)\}$.
 If $(a,\kappa)$ is in $S_3$, the behaviour of the corresponding curve $h(\beta)$ is given
 by Figure \ref{figS3} and there are at most three symmetric solutions of $(GL)$. 

Let us call $\underline{\beta}$ the point where $h(\underline{\beta})=
\underline{h}$. Then as $(a,\kappa)$ tends to a point
 $(\tilde{a},\kappa_2(\tilde{a}))$ on the curve $\kappa_2(a)$, $(\underline{\beta},\underline{h})$
 tends to $(0,h_s)$. As $(a,\kappa)$ tends to a point
 $(\tilde{a},\kappa_3(\tilde{a}))$ on the curve $\kappa_3(a)$, $(\underline{\beta},\underline{h})$
 and $(\overline{\beta},\overline{h})$ have the same limit $(\tilde{\beta},
h(\tilde{\beta)})$
 with $\tilde{\beta}\in (0,1)$. Moreover, $\tilde{\beta}=0$ if and only if $\tilde{a}=a_*$. 
\end{itemize}
\end{conj}

Note that in $S_2$, $h_s$ is sometimes called the supercooling field
 and in $S_2$ and $S_3$, $\overline{h}$ refers to the superheating field. We will
 come back to this in Section 5, since this is linked to local stability.

\noindent
Our physical interpretation of the 3 regions is the following:

\noindent
* $S_1$ corresponds to what is usually called thin films: when the
 size of the slab is sufficiently small, there is no difference between type I and type II
 superconductors.
 The Meissner effect is observed, that is the magnetic field is expelled
 from the material and there is no superheating nor supercooling phenomenon. We refer to \cite{SJ} for more details. In particular, our curves $\kappa_1(a)$ and $\kappa_3(a)$ give a quantitative estimate
 of how small $a$ should be depending on $\kappa$ so that only one dimensionnal phenomena
 happen. 

\hfill

\noindent
* We claim that the curve $\kappa_2(a)$ for $a\geq a_*$ can be considered as the separating curve
 between type I and type II behaviours. 
 Indeed, it is known (see Tinkham \cite{T} for instance for details)
 that when $h_0$ is decreased from infinity, type I materials become superconducting through a first order phase transition and type II materials through a second order phase transition. 
 More precisely, type I materials supercool, that is remain normal until $h_0=h_s$, where nucleation occurs, followed by a discontinuous and irreversible jump in the maximum of $f$. This is due to the fact that the transition between the normal
 phase and the superconducting phase is accompanied by positive energy (the superconducting solution has a positive energy, that is bigger than the energy of the normal solution),
 which corresponds to a bifurcation curve of locally unstable solutions (subcritical bifurcation). On the contrary, for type II materials, $f$ varies continuously at the transition because the superconducting solution has a negative energy, that is lower than the energy of the normal solution, so that the bifurcated curve of superconducting
 solutions is locally stable. Therefore, we can take as 
 a definition of type I (resp. type II) behaviours the fact that the
 bifurcation from the normal solution is subcritical (resp. supercritical).
 Hence, the curve $\kappa_2(a)$ separates the two regimes for finite $a\geq a_*$
 and we have $\lim_{a\to\infty} \kappa_2(a)
=1/\sqrt 2$: region $S_2$ corresponds to type I behaviours, where the bifurcation is subcritical, and region $S_3$ to type II, where the bifurcation is supercritical.
 In the limiting case $a=\infty$, the formal computations
 of Chapman \cite{chapmanI} indicate that the bifurcation curve of symmetric solutions changes from subcritical to supercritical as $\kappa$ passes through
 $1/\sqrt 2$ from below.

\hfill

Below, we give further discussions and interpretations of the regions and the 
 curves depicted in Figure \ref{figkappaasym}.

\subsection{$\kappa_1(a)$}

For $\kappa=0$, Ginzburg \cite{G} obtained the critical value $a_0=\sqrt 5 /2$:
 if $a<a_0$, $h(\beta)$ is decreasing (Figure \ref{figS1})
 so that no supercooling nor superheating is possible. For $a>a_0$, $h(\beta)$ has the behaviour described by Figure \ref{figS2}
 and displays both supercooling and superheating phenomena.

Our study indicates that as $(a,\kappa)$ passes through $\kappa_1(a)$,
 the bifurcation at $h_s$ changes from supercritical in $S_1$
 to subcritical in $S_2$. Along the curve $\kappa_1(a)$, the function $h(\beta)$
 is decreasing and we believe that $h''(0)=0$, or equivalently,
 the term $h_1=0$ in the development of $h$ in (\ref{bifurcatedsol}). This means that as $(a,\kappa)$ enters $S_2$, the fold in the $h(\beta)$ curve is going to appear
 near $\beta=0$. The fact that $h''(0)=0$ on the curve $\kappa_1(a)$
 is directly linked to Ginzburg's observation that at the transition, the specific heat (which can be related to the inverse of  a derivative of the energy) discontinuity  
 is infinite.
 Further insight into the nature of this bifurcation could be 
 obtained if one were to analyze a formula indicating the direction of bifurcation. Such a computation is directly linked to the evaluation of the term $h_1$
 in (\ref{bifurcatedsol}). We refer to  
 Millman-Keller \cite {MK} for a formula for $h_1$. 

\subsection{$\kappa_2(a)$}

This is the limiting curve separating type I and type II behaviours as we have explained earlier.
 Our study indicates that as $(a,\kappa)$ passes from $S_2$ to $S_3$ across $\kappa_2(a)$, there is
 a change from  subcritical bifurcation in $S_2$ to supercritical in $S_3$.
 Moreover, along the curve $\kappa_2(a)$,
 for $a>a_*$ the function $h(\beta)$ has the graph indicated in Figure \ref{figS2}. Additionnally $h''(0)=0$, since $h'(\beta)$ changes sign 
 near $\beta=0$. This is similar to what is happening on $\kappa_1(a)$. Again by computing $h_1$ exactly as described above for $\kappa_1(a)$, one can obtain further mathematical results about the nature of the bifurcation
 and the location of $\kappa_2(a)$. 

The following additional information has been obtained in \cite{BH8} about the asymptotic behaviour of the curve when $a$ is large:
\begin{theo}\label{kappalargeBHS3} (Bolley-Helffer \cite{BH8})
There exists a constant $\kappa_0(a)>1/ \sqrt 2$ such that for
 $\kappa \geq \kappa_0(a)$
  and for $h_0$ the eigenvalue of a spectral problem, then the
 bifurcated solution (\ref{bifurcatedsol}) bifurcating from the normal solution
 $(0,h_0x,h_0)$ satisfies $h(\eps)<h_0$. Moreover, there exists a constant
 $C$ such that $\kappa_0(a)$ can be chosen such that
$${1 \over {\sqrt 2}} <\kappa_0(a)= {1 \over {\sqrt 2}}+O(exp(-Ca^2))\quad \hbox{when}\quad a \to \infty.$$
\end{theo}

\subsection{$\kappa_3(a)$}
\begin{figure}[htb]
$$\epsfxsize=9cm
\epsfysize=5.5cm
\epsfbox{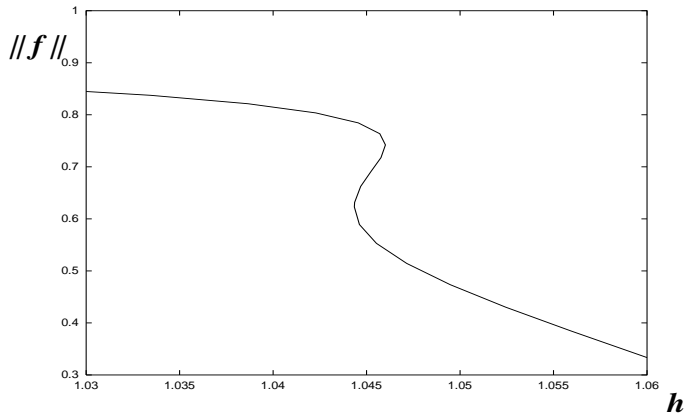}$$
\vspace{-1.2cm}
\caption{Part of the curve $h(\beta)$ for $a=2$ and $\kappa=1$.}
\label{figS3zoom}
\end{figure}

Our study indicates that as $(a,\kappa)$ passes across $\kappa_3(a)$,
 the two folds at $(\overline{\beta},\overline{h})$ and $(\underline{\beta},\underline{h})$ which exist for a solution in $S_3$ tend to the same limit
 $(\tilde{\beta},\tilde{h})$. As a result,
 the function $h(\beta)$ is decreasing along $\kappa_3(a)$,
 but there is an interior point $\tilde{\beta}$ with infinite slope, that is $h''(\tilde{\beta})=0$: this is
where the curve gains its S shape as $(a,\kappa)$ enters $S_3$ (see Figure \ref{figS3zoom}).
 At the left end of $\kappa_3(a)$, that is at the triple point
 $(a_*,\kappa_*)$, we have $\tilde{\beta}=0$.
 It is an interesting open mathematical problem to define the curve $\kappa_3(a)$ by means of a spectral problem in terms of $h''(\beta)=0$ for some $\beta\in [0,1)$.



\subsection{The triple point}
As described earlier, our numerical experiments indicate the existence of
 a unique point $a_*\simeq 1.60$ for which $\kappa_1(a_*)=\kappa_2(a_*)=
\kappa_3(a_*)=\kappa_*$. Thus, in Figure \ref{figkappaasym} we see that this `triple point' $(a_*,\kappa_*)$  provides the common intersection of the three
 curves. To our knowledge, the existence of this important point
 has not previously been reported in the literature.
  At the point itself, we observe that 
 $h_s=h_*\simeq 1.2$, $h(\beta)$ decreases on $(0,1)$ and we believe that 
 $h''(0)=h^{(iv)}(0)=0$, that is $h_1=h_2=0$ in (\ref{bifurcatedsol}).
 However, if we let $(a,\kappa)$ vary from the triple point along rays leading
 into the interior of $S_1$, $S_2$ and $S_3$, we expect to see entirely different behaviours. 
 Along a ray leading into $S_1$ from the triple point, the bifurcation diagram for symmetric
 solutions is the same as that shown in Figure \ref{figS1}. Similarly, along rays leading into
 $S_2$ and $S_3$, the bifurcation diagrams for symmetric solutions are the same as those
 shown in Figure \ref{figS2} and \ref{figS3}, respectively.

Clearly, a careful bifurcation analysis needs to be done to 
 expose the behaviour of solutions in a neighbourhood of the point
 $a=a_*$, $\kappa=\kappa_*$, $h=h_*$.


\subsection{Region $S_1$}
Recall that the behaviour of $h(\beta)$ is described by Figure \ref{figS1}.
\subsubsection{$\kappa$ small}

This regime was originally studied by Ginzburg \cite{G} in the case $\kappa=0$.
 He uses
 an approximate model in which $f$ is held constant, i.e. $f\equiv\beta$.
 With this approximation, he observes
 that for $a<\sqrt 5/2$, the curve $h(\beta)$ is decreasing, as described in
 Figure \ref{figS1}. He computes $h_s=\sqrt 3/a$ and claims that the approximate model remains valid 
 in describing the properties of the solutions for small $\kappa$. 
 
For the approximate model studied by Ginzburg, his computations have been made rigorous by Bolley-Helffer \cite{BH2}-\cite{BH3}.

For the full system $(GL)$, there is only the following local result concerning
 the bifurcation
 of symmetric solutions from the normal solution. 
\begin{theo}\label{kappaasmallBH} (Bolley-Helffer \cite{BH8})
For any $\eta>0$, there exists $C_0$ such that for $\kappa a\leq C_0$,
 $\sqrt 5/2 -a>\eta$ and $h_0$ the eigenvalue of a spectral problem, then the
 bifurcated solution (\ref{bifurcatedsol}) starting from the normal solution
 $(0,h_0x,h_0)$ satisfies $h(\eps)<h_0$. In particular, the bifurcation is supercritical.
\end{theo}
Rigorous treatment of the global problem remains open.
\subsubsection{$\kappa$ large}

Recently, this regime has received an increasing amount
 of attention since high temperature superconductors
 typically have large $\kappa$. For mathematical results, see Berestycki, Bonnet,
 Chapman \cite{bebecha}
 or  Aftalion \cite{aa1}.

\begin{conj}\label{klarge}
Let $a$ be fixed. When $\kappa$ tends to $\infty$,
 the limiting behaviour of the curve $h(\beta)$ is the union
 of a horizontal line at $\beta=1$ from $h=0$ to $h_0^*(a)$,
 and a curve joining $(h_0^*(a),1)$
 to $(\kappa,0)$, where $h_0^*(a)$ is given by
$$a=\int_{0}^{1} du\ {\Bigl ({h_0^*}^2-{1\over 2}
(1-u^2)^2\Bigr )}^{-{1 \over 2}}.$$
\end{conj}
As explained in \cite{aa1}, the limiting problem when $\kappa$ tends to
 $\infty$ is
$$\left\{\begin{array}{ll}
f^2=(1-q^2)\hbox{\bf 1}_{|q|\leq 1}
\quad\hbox{in}\quad (0,a),\\
q''=q(1-q^2)\hbox{\bf 1}_{|q|\leq 1} \quad\hbox{in}\quad (0,a),\\
q'(\pm a)=h_0.
\end{array}\right.\eqno{(GL_{\infty})}$$
It is proved in \cite{aa1} that there is a unique solution $(f_\infty,q_\infty)$ of $(GL_\infty)$
 for $h_0\leq h_0^*$ and $|q_\infty|\leq 1$ so that $|f_\infty|>0$. This solution is symmetric, i.e. $q(0)=0$ and $f(0)=1$. The critical field $h_0^*$ is determined by the extra condition $q(a)=1$.
 Moreover,  any minimizer of $E_\kappa$ converges to $(f_\infty,q_\infty)$
 as $\kappa$ tends to $\infty$. Thus, it is natural to expect that the minimizer
 for finite $\kappa$ is symmetric and gives the horizontal line of the
 $h(\beta)$ diagram.

Note that $\lim_{a\to\infty} h_0^*(a)=1 / \sqrt 2$, which will be the same
 limit later in $S_3$ for $a=\infty$, $\kappa$ large.

Further study is needed to resolve the behaviour of solutions in this regime.



\subsection{Region $S_2$}
Recall that the behaviour of $h(\beta)$ is described by Figure \ref{figS2}.
\subsubsection{$\kappa$ small}

This regime was also studied by Ginzburg \cite{G} with the approximate model where $f$ is constant. He observes
 that for $a>\sqrt 5/2$, the curve $h(\beta)$ has the same behaviour as the one described by
 Figure \ref{figS2}. 
 Again, only the computations for the approximate model have been made rigorous by Bolley-Helffer \cite{BH4}. In particular, they prove that, for small $\kappa$, the curve $h(\beta)$ has a local maximum $\overline{h}$, for which they give 
 asymptotic formula consistent with physicists'results.

The only result to date for the full system $(GL)$ concerns the bifurcation
 of symmetric solutions from the normal solution: 
\begin{theo}\label{kappaasmallBHS2} (Bolley-Helffer \cite{BH8})
For any $\eta>0$, there exists $C_0$ such that for $\kappa a\leq C_0$,
 $a-\sqrt 5/2 >\eta$ and $h_0$ the eigenvalue of a spectral problem, then the
 bifurcated solution (\ref{bifurcatedsol}) starting from the normal solution
 $(0,h_0x,h_0)$ satisfies $h(\eps)>h_0$. In particular, the bifurcation is subcritical.
\end{theo}

\subsubsection{$a$ large}
Here the first rigorous multiplicity result is the following:
\begin{theo}\label{HKTS2} (Hastings, Kwong $\&$ Troy, \cite{HKT})
 Let $\kappa\leq 1/\sqrt 2$ be fixed. Then for a large enough, there is a $\beta_1$
 in $(1-1/(\kappa a),1)$ such that 
 $h(\beta_1)>h_s$. In particular, this means that $S_2 \neq \emptyset$ and that there are values of $h_0$
 such that $(GL)$ has two symmetric solutions.
\end{theo}
From  \cite{HKT}, we also have the following limit, which was originally
 predicted by Ginzburg-Landau \cite{GL}:
\begin{prop}\label{hsinfty} (Hastings, Kwong $\&$ Troy, \cite{HKT})
$\lim_{a\to \infty}h_s=\kappa$.
\end{prop}
The only bifurcation result available at the moment is the following:
\begin{theo}\label{kappaalargeBHS2} (Bolley-Helffer \cite{BH8})
For any $\eta>0$, there exists $C_0$ such that for $\kappa a\geq C_0$,
 $1/ \sqrt2 -\kappa>\eta$ and $h_0$ the eigenvalue of a spectral problem, then the
 bifurcated solution (\ref{bifurcatedsol}) starting from the normal solution
 $(0,h_0x,h_0)$ satisfies $h(\eps)>h_0$. In particular, the bifurcation is subcritical.
\end{theo}

For the limiting behaviour of the curve $h(\beta)$
 when $a$ is large, our experiments lead us to the following.
\begin{conj}\label{ainfty}
Let $\kappa \leq 1/ \sqrt 2$ be fixed. When $a$ tends to $\infty$,
 the limiting behaviour of the curve $h(\beta)$ is the union
 of a horizontal line at $\beta=1$ from $h=0$ to $\overline{h}_\infty>1/\sqrt 2$,
 and a curve joining $(h=1/ \sqrt 2,\beta=1)$
 to $(h=\kappa,\beta=0)$.
\end{conj}
The schematic limiting shape of the diagram can be found in Figure \ref{figainfk.3} or \ref{figainfk.5} at the end of Section 4. The curve `Sym' corresponds to symmetric solutions.
 It means that the fold is getting close to the line $\beta=1$ and the knee
 is getting very thin, as illustrated by Figure \ref{figa10k.3} in Section 4, where the extra
 curve corresponding to asymmetric solutions has been added.
 To help resolve this conjecture, one should investigate two relevant boundary value problems which we now describe: the first one is for
 the curve from $(1/ \sqrt2,1)$ to $(\kappa,0)$ in Figure \ref{figainfk.5}. It is derived by letting $a$ tend to $\infty$ in $(GL_{sym})$:
\beq\label{S2sym1}\left\{\begin{array}{ll}
{1 \over \kappa^2}f''=f(f^2+q^2-1)
\quad\hbox{in}\quad (0,\infty),\\
f(0)=\beta,\quad f'(0)=0,\\
q''=qf^2 \quad\hbox{in}\quad (0,\infty),\\
q(0)=0,\quad q'(0)=\alpha,
\end{array}\right.\eeq
\begin{conj}\label{conjS2infty}
Let $\kappa< 1/ \sqrt 2$ be fixed. For each $\beta$ in $(0,1)$, there exists
 a unique $\alpha$ such that the solution of (\ref{S2sym1}) satisfies $\lim_{x\to\infty}f(x)=0$. 
 We define $h(\beta)=\lim_{x\to\infty}q'(x)$. Then $h(\beta)$ is a continuous increasing
 function of $\beta$, 
\beq\label{S2hinfty}
\lim_{\beta\to 0} h(\beta)=\kappa \quad \hbox{and}
\quad \lim_{\beta\to 1} h(\beta)={1\over \sqrt 2}. \eeq
\end{conj}

\begin{theo}\label{1sqrt2}
Let $\kappa=1/\sqrt 2$. For each $\beta$ in $(0,1)$, there exists
 an $\alpha$ such that the solution of (\ref{S2sym1}) satisfies $\lim_{x\to\infty}f(x)=0$. Moreover,
 $$\lim_{x\to\infty}q'(x)=1/\sqrt 2 \quad \forall \beta\in (0,1).$$
\end{theo}
This is proved by Chapman in \cite{chapmanI} using the special decoupling of $(GL)$ for $\kappa=1/\sqrt 2$ mentionned
 in \cite{CMLHO}:
\beq
\left\{\begin{array}{ll}
S=f'+fq/\sqrt 2,\\
F=q'+(f^2-1)/\sqrt 2.
\end{array}\right.
\eeq
Chapman observes in \cite{chapmanI} that a symmetric solution of $(GL)$ leads
 to $(S(0),$ $F(0))=(0,0)$. It follows from uniqueness of solutions of a first order system that $(S,F)\equiv (0,0)$
 and therefore the full $(GL)$ system reduces to 
\beq
\left\{\begin{array}{ll}
f'=-fq/ \sqrt 2,\\
q'=(1-f^2)/ \sqrt 2.
\end{array}\right.
\eeq

\hfill

The second BVP relevant to Conjecture \ref{ainfty} corresponds to 
 the horizontal line $\beta=1$ in the bifurcation diagram of Figure \ref{figainfk.5}. 
 It comes from a boundary layer analysis: the extremal part of the slab
 is sent to 0 and the symmetry plane to $\infty$, so that we have 
\beq\label{S2sym2}\left\{\begin{array}{ll}
{1 \over \kappa^2}f''=f(f^2+q^2-1)
\quad\hbox{in}\quad (0,\infty),\\
f'(0)=0,\quad \lim_{x\to\infty}f(x)=1,\\\
q''=qf^2 \quad\hbox{in}\quad (0,\infty),\\
q'(0)=h,\quad \lim_{x\to\infty}q(x)=0.
\end{array}\right.\eeq

This problem and the boundary layer analysis were originally derived by Ginzburg \cite{G}, and \cite[Figure 5]{G} leads to the following conjecture:
 
\begin{conj}\label{conj2S2infty}
Let $\kappa>0$ be fixed. There exists $\overline{h}_\infty>1/\sqrt 2 $
 such that 
 (\ref{S2sym2}) has a unique solution
 for $h\leq1/\sqrt 2$ and two solutions for $1/\sqrt 2<h<\overline{h}_\infty$.
\end{conj}
This explains why the curve going down from $\beta=1$ to $\beta=0$ in Figure \ref{figainfk.5} starts at
 $h=1 / \sqrt 2$.
 For this conjecture, partial results have been provided
 by Bolley-Helffer:
\begin{theo}\label{superheatingBH}
(Bolley-Helffer \cite{BH7})
For any $\kappa>0$, there exists $\overline{h}_\infty (\kappa)$ such that
 there is no solution of (\ref{S2sym2}) if $h>\overline{h}_\infty (\kappa)$, and at least a solution if $h<\overline{h}_\infty$.
 Moreover,
\beq\label{degennes}
\lim_{\kappa\to 0}\kappa\overline{h}_\infty ^2(\kappa)={{\sqrt 2}\over 4}.
\eeq
\end{theo}
This formula was first derived by De Gennes.

\subsection{Region $S_3$}
Recall that the behaviour of $h(\beta)$ is described by Figure \ref{figS3}.
\subsubsection{$a$ large}
Here the only rigorous multiplicity result is the following:

\begin{theo}\label{HKTS3} (Hastings, Kwong $\&$ Troy, \cite{HKT})
Let $\kappa>1/\sqrt 2$ be fixed. Then for a large enough, there are $\beta_1$
 and $\beta_2$ in $(1-1/(\kappa a),1)$ such that $\beta_2>\beta_1$,
 $h(\beta_1)<h_s$ and $h(\beta_2)>h(\beta_1)$. In particular, this means that
 the set $S_3 \neq \emptyset$ and that there are
 values of $h_0$ such that $(GL)$ has at least 3 solutions.
\end{theo}

\begin{conj}\label{ainftyS3}
Let $\kappa > 1/ \sqrt 2$ be fixed. When $a$ tends to $\infty$,
 the limiting behaviour of the curve $h(\beta)$ is the union
 of a horizontal line at $\beta=1$ from $h=0$ to $\overline{h}_\infty>1/\sqrt 2$,
 and a curve joining $(h=1/ \sqrt 2,\beta=1)$
 to $(h=\kappa,\beta=0)$.
\end{conj}
Notice that the bifurcation curve for $a$ large has the same limiting
 shape as in $S_2$, but this time the curve from $(1/ \sqrt2 ,1)$ to
 $(\kappa,0)$ has negative slope. 
 This is illustrated by Figure \ref{figainfk.9} below with the curve `Sym' corresponding
 to symmetric solutions. The part of the diagram corresponding
 to the line $\beta=1$ is described by the same limiting problem as in $S_2$.
 Thus, the field $\overline{h}_\infty$ is the value derived from Conjecture \ref{conj2S2infty}.

\begin{conj}\label{conjS3infty}
Let $\kappa> 1/ \sqrt 2$ be fixed. For each $\beta$ in $(0,1)$, there exists
 a unique $\alpha$ such that the solution of (\ref{S2sym1}) satisfies $\lim_{x\to\infty}f'(x)=0$.
 We define $h(\beta)=\lim_{x\to\infty}q'(x)$. Then $h(\beta)$ is a
 continuous decreasing
 function of $\beta$, and
\beq\label{S3hinfty}
\lim_{\beta\to 0} h(\beta)=\kappa \quad \hbox{and}
\quad \lim_{\beta\to 1} h(\beta)={1\over \sqrt 2}. \eeq
\end{conj}

\subsubsection{$\kappa$ large, $a=\infty$}
The asymptotic behaviour of the upper bound for the superheating field 
  $\overline{h}_\infty (\kappa)$ in $S_3$ has been proved in the following:
\begin{theo}\label{superheatingBHS3}
(Bolley-Helffer \cite{BH7})
The field $\overline{h}_\infty (\kappa)$ defined in Theorem \ref{superheatingBH} satisfies
\beq\label{hinfty}
\lim_{\kappa\to \infty}\overline{h}_\infty (\kappa)={1\over {\sqrt 2}}.
\eeq
\end{theo}
This proves part of the formal asymptotic development derived by Chapman \cite{chap2}: $\overline{h}_\infty=1/ \sqrt 2+0.33 \kappa^{-4 /3}+o(\kappa^{-4 /3})$
 as $\kappa$ tends to $\infty$. Notice that this means that
 the `superheating part' of the curve is getting small (recall Conjecture \ref{conjS3infty}) and is consistent with the fact that we reach the boundary
 of $S_1$. We find the same limit as $\lim_{a\to\infty}h_0^*(a)$.

\subsubsection{$\kappa a$ large}
The only result so far concerning the bifurcation of symmetric solutions
 in this regime is the following
\begin{theo}\label{kappaalargeBHS3} (Bolley-Helffer \cite{BH8})
For any $\eta>0$, there exists $C_0$ such that for $\kappa a\geq C_0$,
 $\kappa -1/ \sqrt2 >\eta$ and $h_0$ the eigenvalue of a spectral problem, then the
 bifurcated solution (\ref{bifurcatedsol}) starting from the normal solution
 $(0,h_0x,h_0)$ satisfies $h(\eps)<h_0$. In particular, the bifurcation is supercritical.
\end{theo}


\section{Asymmetric solutions}
\setcounter{equation}{0}

\begin{figure}[p]
\vspace{-1.5cm}
$$\epsfxsize=9cm
\epsfysize=5.5cm
\epsfbox{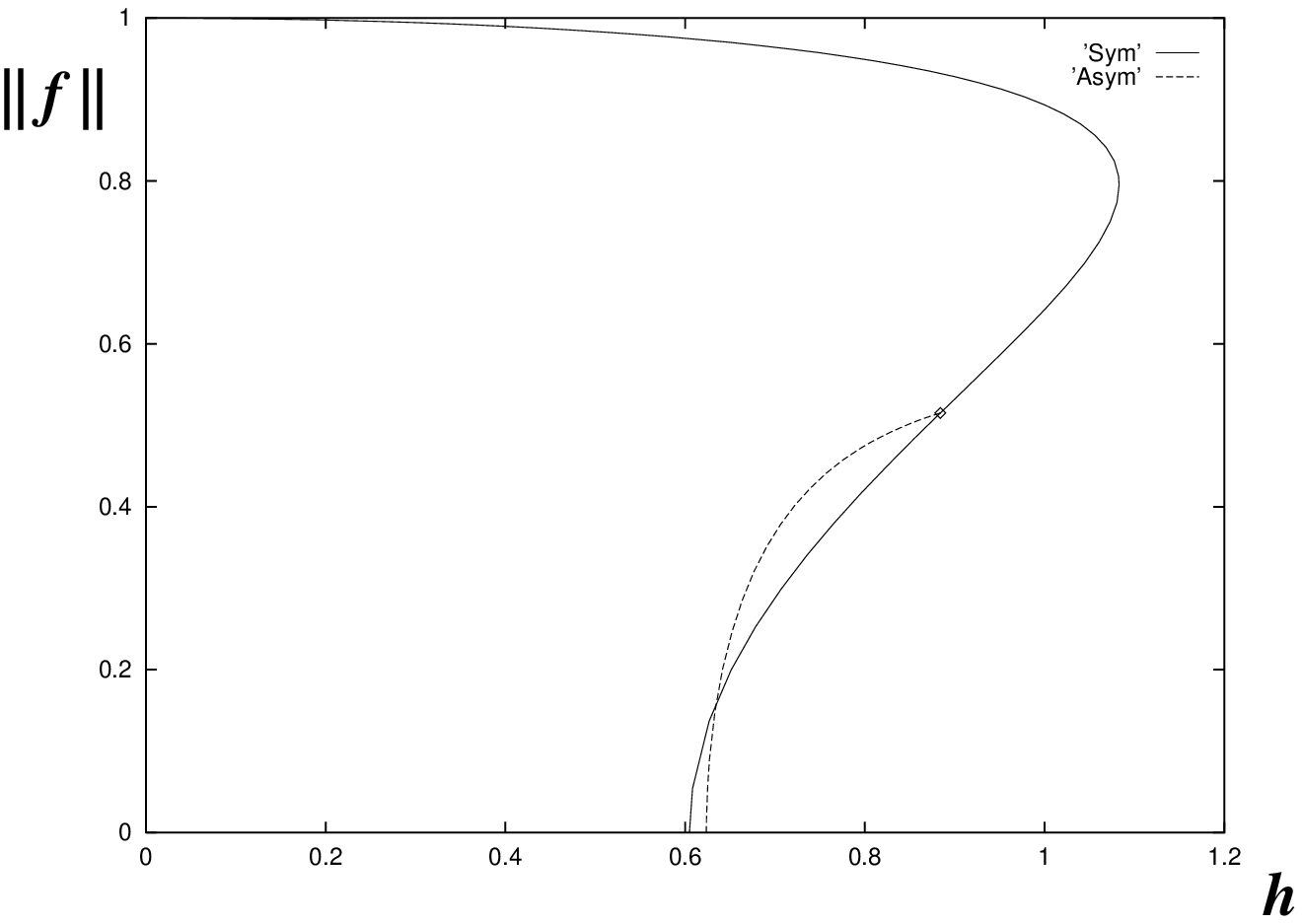}$$
\vspace{-1.2cm}
\caption{Bifurcation diagrams for $a=3$ and $\kappa=0.35$.}
\label{figa3k.35}
$$\epsfxsize=9cm
\epsfysize=5.5cm
\epsfbox{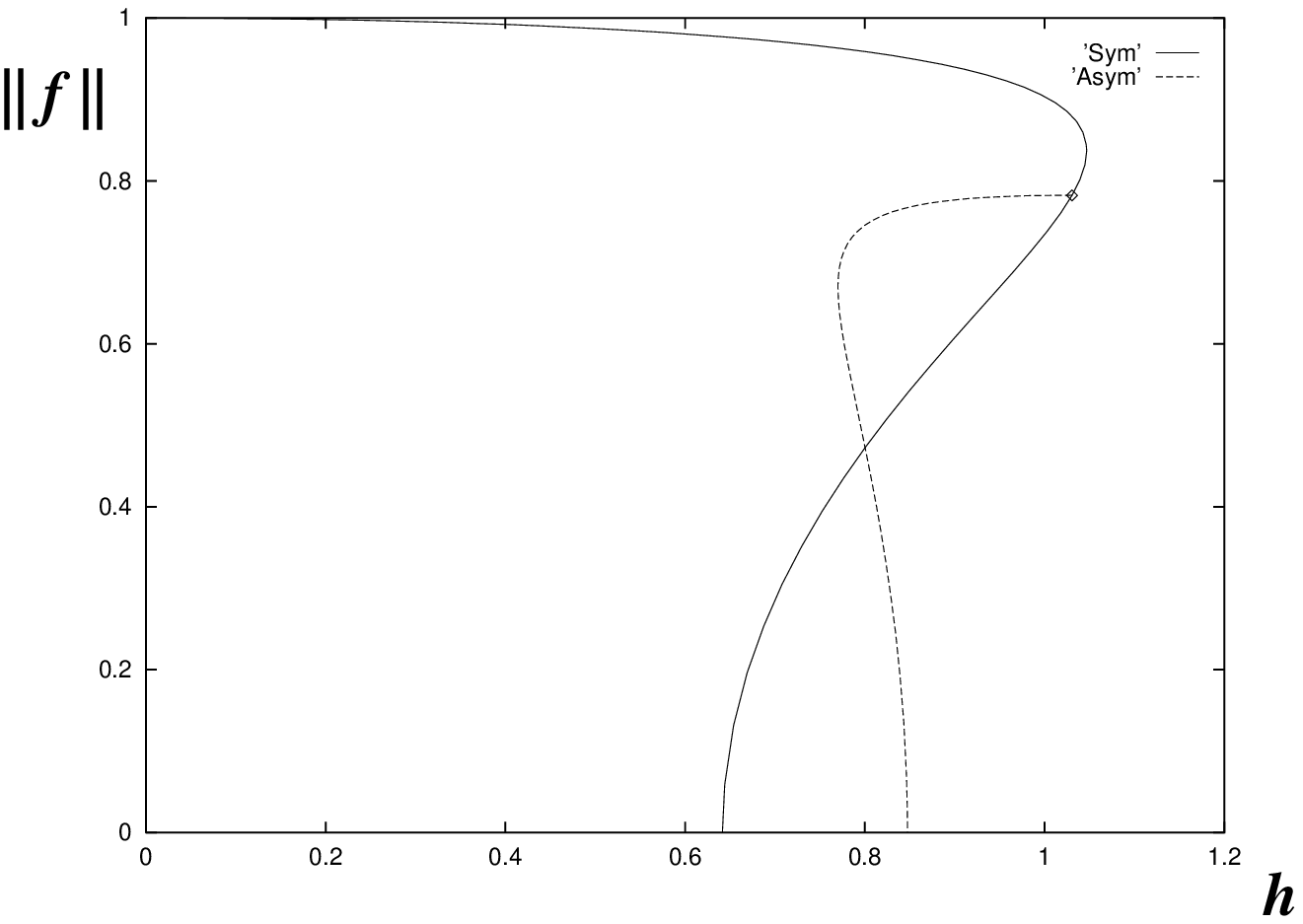}$$
\vspace{-1.2cm}
\caption{Bifurcation diagrams for $a=3$ and $\kappa=0.5$.}
\label{figa3}
\begin{center}
\input autoAkappaa.pstex_t
\end{center}
\vspace{-1cm}
\caption{Curves $\kappa_4(a)$ and $\kappa_5(a)$.}
\label{figkappaaasym}
\end{figure}

Recall from Section 2 that $(f,q)$ is an asymmetric solution
 of $(GL)$ if it is not symmetric. Notice that if $(f(x),$ $ q(x))$
 is a solution, then so is $(f(-x), q(-x))$, but in the following
 we will count this as one solution instead of two and assume that the maximum  of $f$ is reached in $[-a,0]$.
 Below, we summarize the results of our numerical investigation of asymmetric solutions shown in Figures \ref{figa3k.35} and \ref{figa3}: the vertical axis has been labeled $\|f\|$. For the symmetric branch, $\|f\|$ refers to $f(0)=\beta$ as in Figures \ref{figS1}, \ref{figS2} and \ref{figS3}.
 For the asymmetric branch, $\|f\|=\|f\|_\infty$, which is attained 
 at some point $x_\beta$ different from zero. For simplicity, we will sometimes
 refer to the complete bifurcation diagram as $h(\beta)$.
  In Figures \ref{figa3k.35} and \ref{figa3}, the curve
 $h(\beta)$ for symmetric solutions is the typical curve in $S_2$. On these curves, there
 is a point $(\beta_b,h_b)$ denoted by a square: it is the branching point of the branch of asymmetric solutions.
 More precisely, for  $\|f\|_\infty$ close to $\beta_b$, the asymmetric solution is nearly symmetric.
 The branch of asymmetric solutions leads from the branch of symmetric solutions 
 to the branch of normal solutions. If the branch of asymmetric solutions crosses the branch
 of symmetric solutions without any square indicated, it just means that for this special $h_0$,
 $(GL)$ has a symmetric solution and an asymmetric solution having the same maximum value
 of $f$. We have observed the following:
\begin{conj}\label{xbeta}
Let $x_\beta$ be the point where $f$ reaches its maximum. Then $x_\beta$ is
 a decreasing function of $\beta$ and
 $ \lim_{\beta \to \beta_b} x_\beta=0.$
\end{conj}
We now describe the two possible behaviours of the {\em asymmetric curve} $h(\beta)$.
\begin{itemize}
\item Figure \ref{figa3k.35}: for $\beta\in (0,\beta_b)$, $h(\beta)$ is an increasing
 function of $\beta$ and $\lim_{\beta\to 0} h(\beta)= h_{as}$,
 $\lim_{\beta\to \beta_b} h(\beta)= h_b$, with $h_s<h_{as}<h_b\leq \overline{h}$.
 If $h_0 \in (h_{as}, h_b)$, there is a unique asymmetric solution of $(GL)$ and if $h_0\leq h_{as}$
 or $h_0\geq h_b$, no such solution.

\item Figure \ref{figa3}: for $\beta\in (0,\beta_b)$, $h(\beta)$ is a decreasing
 function of $\beta$  until $h$ reaches a minimum value $\underline{h}_{as}$ and then
 increasing to $h_b$ as $\beta$ goes to $\beta_b$. thus, 
  for $h_0 \in (\underline{h}_{as}, h_b)$, there are two asymmetric solution of $(GL)$,
 for $h_0\in (h_b,h_{as})$, a unique asymmetric solution and for $h_0< \underline{h}_{as}$
 or $h_0\geq h_b$, no such solution. 
\end{itemize}
This implies in particular that for Figure \ref{figa3k.35} the bifurcation of asymmetric solutions
 from the normal solution is subcritical 
 and for Figure \ref{figa3} supercritical.

 In Figure \ref{figkappaaasym}, we have computed two new curves, $\kappa_4(a)$ and $\kappa_5(a)$ which play an
 essential role in the study of  existence and multiplicity of asymmetric solutions. 

\begin{conj}\label{asym}
There exist two continuous functions  $\kappa_4(a)$ and $\kappa_5(a)$
 separating the $(a,\kappa)$ plane into three
 connected regions
 $A_0$, $A_1$ and $A_2$. 
 There exists exactly one point $a^*$  called the quintuple point
 (with approximate value 1.23) such that $\kappa_4(a^*)=\kappa_5(a^*)
=1/\sqrt 2$ and $1/ \sqrt 2 =\kappa_1(a^*)$.
 Moreover, $\kappa_4(a)=C/a$ with C approximately equal to 0.90,
 and
 $\kappa_5(a)$ is defined and monotone decreasing on
  $[a^*,\infty)$ with $\lim_{a\to \infty} \kappa_5(a)=\kappa_{as}\simeq 0.4$.
\begin{itemize}
\item region $A_0=\{a>0,0<\kappa\leq \kappa_4(a)\}$. There are
 no asymmetric solutions of $(GL)$. Thus in the $h(\beta)$ plane, we only see curves as shown in Figure \ref{figS1} and \ref{figS2}.
\item region $A_1=\{a^*< a,\kappa_4(a)<\kappa\leq \kappa_5(a)\}$. 
 If $(a,\kappa)$ is in $A_1$, the bifurcation diagram for asymmetric solutions is given by
 Figure \ref{figa3k.35}, it has no fold and there is at most one asymmetric solution of  $(GL)$.

 Moreover, when $(a,\kappa)$ tends a point $(\tilde{a},\kappa_4(\tilde{a}))$ on the curve
 $\kappa_4(a)$,
 then $h_{as}$ and $h_b$ tend to $h_s$. When $(a,\kappa)$ tends to some
 $(\tilde{a},\kappa_5(\tilde{a}))$,
 then $h_{as}$ tends to $h_s$.
 
\item region $A_2=\{0<a\leq a^*,\kappa> \kappa_4(a)\}\cup\{a>a^*,\kappa>\kappa_5(a)\}$.
  If $(a,\kappa)$ is in $A_2$, the bifurcation diagram for asymmetric solutions is given by
 Figure \ref{figa3}, it has a unique fold and there are at most two asymmetric solutions of $(GL)$.

Moreover, when $(a,\kappa)$ tends to a point  $(\tilde{a},\kappa_4(\tilde{a}))$ on $\kappa_4(a)$,
 then $h_{as}$ and $h_b$ tend to $h_s$. When $(a,\kappa)$ tends to a point   $(\tilde{a},\kappa_5(\tilde{a}))$ on $\kappa_5(a)$,
 then $h_{as}$ tends to $h_s$ and $\underline{h}_{as}$ to $h_b$.
\end{itemize}
\end{conj}
Of particular physical interest is the following:
\begin{conj}\label{conjhas}
Let $a>0$. If $\kappa>\kappa_4(a)$ then $h_{as}>h_s$.
\end{conj}
Thus, as $h_0$ decreases from infinity, the material 
 is in the normal state until $h_0$ reaches the
 value $h_{as}$ where the bifurcation of asymmetric solutions occurs.
 If $a>a^*$ and $\kappa >\kappa_5(a)$, or if $a\leq a^*$ and $\kappa>\kappa_4(a)$, then
 the bifurcating asymmetric branch is supercritical and plays an important role in the onset of superconductivity in the material.
 In the following we explain the implications of Conjecture \ref{conjhas}.
 In the physics literature, $h_{as}$ is defined as $h_{c_3}$ and $h_s$ as
 $h_{c_2}$. It is well known to physicists that when $\kappa$ is fixed bigger than $0.4$ and
 $a$ is taken large enough (which corresponds to $\kappa$ being above $\max(\kappa_4(a),\kappa_5(a))$),
 when the exterior magnetic field $h_0$ is decreased from infinity,
 superconductivity is not nucleated first in the volume of the sample,
 which would give rise to symmetric solutions, but rather in a sheath
 near the surface, due to the existence of asymmetric solutions.
 This is called surface superconductivity. 
 If the slab is very thick, the two surface solutions $f(x)$ and $f(-x)$
 do not interact. In this region, superconductivity is first nucleated
 in surface layers of size $1/ \kappa$ near the boundaries, and the middle part of the material is normal. Now, if the slab is of intermediate
 size, the solutions $f(x)$ and $f(-x)$ interfere to create vortices.
 Indeed, recall that the original Ginzburg-Landau energy is gauge invariant
 so that a solution $(f(x),q(x))$ has the same energy as $(e^{i\kappa c y}f(x),q(x)+c)$ for any constant $c$. 
 Thus, when the sample is not too
 large, and $h$ decreases below $h_{as}$, the linear combination
 of the asymmetric solutions $f(x)$ and $f(-x)$ create two dimensionnal
 vortices along the mid-plane $x=0$. This is reflected in the formula:
\beq\label{psi}
\psi=\cos k y (f(x)+f(-x))+i\sin k y (f(x)-f(-x))
\eeq
Further discussion of vortex formation and the details of the derivation of formula (\ref{psi}) are given in Tinkham \cite{T}. 
 In particular, the distance between two vortices along the $x=0$ plane is proportionnal to $1/x_\beta$.
 Indeed, when moving up the asymmetric branch $h(\beta)$, we find that $x_\beta$ tends to zero as $\beta$ approaches the branching point on the symmetric branch. This means that the distance between the two vortices 
 tends to infinity. At the limit, the material is perfectly superconducting because $f$ is symmetric.

 In fact, as described in \cite{TT}, $h_s$
 and $h_{as}$ can both be measured: $h_{as}$ by nucleation of superconductivity
 as we have seen and $h_s$ by magnetic transition, which needs bulk superconductivity to take place.

 Boeck and Chapman \cite{BC} have studied surface superconductivity in detail
 and they have determined
 the regimes of $(a,\kappa)$ where the asymmetric solution gives rise to a surface sheath or to vortex solutions, through formula (\ref{psi}).

Now recall that the surface sheath is of size $1/\kappa$ (that is of order of the coherence length $\xi$), so it is consistent that a condition for surface
 superconductivity to exist should be $a\geq C / \kappa$. As mentionned by Tinkham \cite[p.136]{T}, physicists had found that $2C\simeq 1.81$ but did not know how to predict the details of changeover of behaviour on this curve.
 We believe that our numerical results answer this open question. A curve similar to $\kappa_4(a)$ was also mentionned
 by Boeck and Chapman \cite{BC}. 

\hfill

The global results of our numerical investigations are shown graphically in Figure
 \ref{figsummary} where for each region of interest, we refer to the relevant Figure describing the bifurcation
 diagram of both symmetric and asymmetric solutions.
\begin{figure}[htb]
\begin{center}
\input autokafig.pstex_t
\end{center}
\vspace{-1cm}
\caption{Curves $\kappa_1(a)$, $\kappa_2(a)$, $\kappa_3(a)$, $\kappa_4(a)$ and $\kappa_5(a)$.}
\label{figsummary}
\end{figure}
\begin{figure}[htb]
$$\epsfxsize=9cm
\epsfysize=5.5cm
\epsfbox{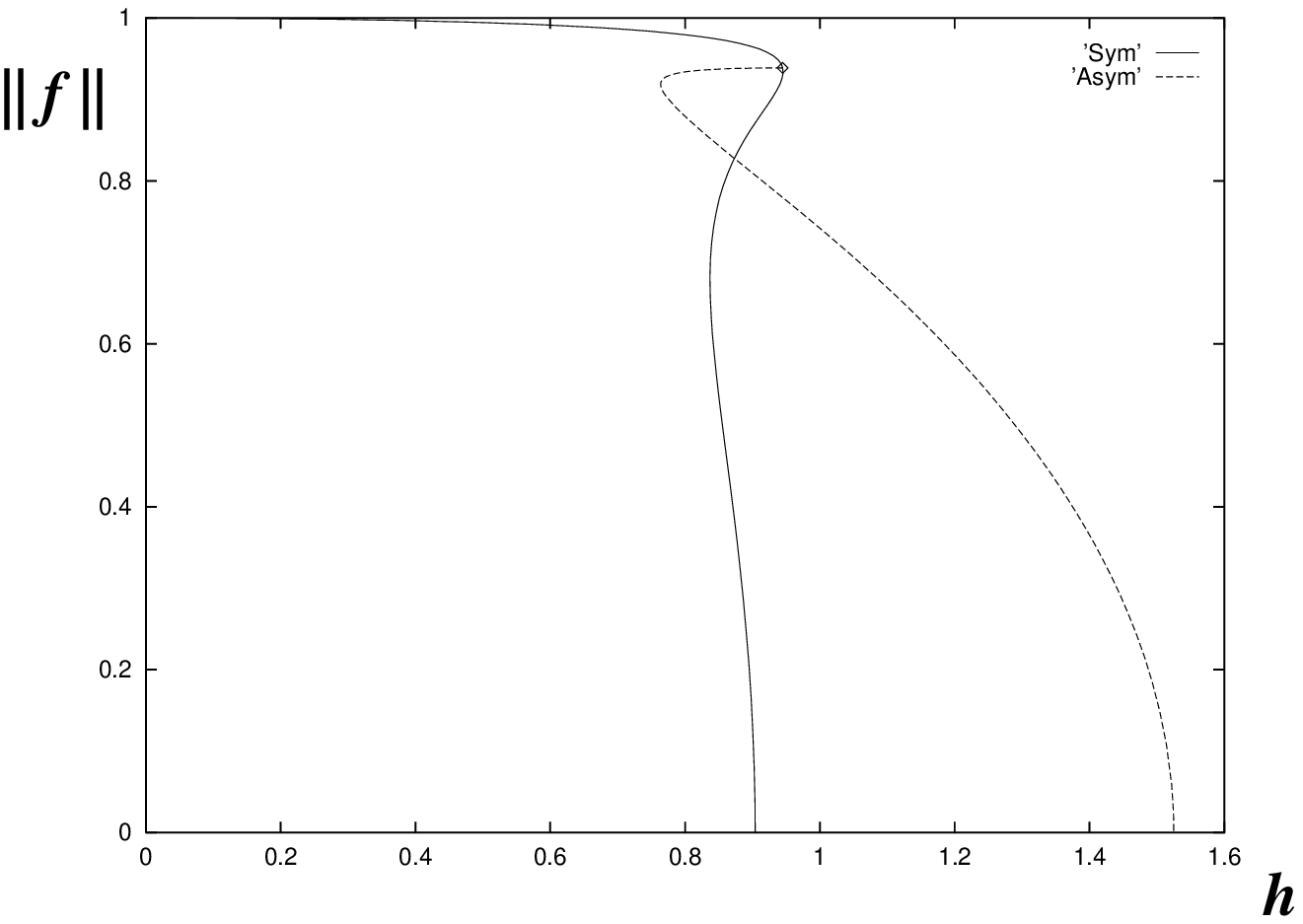}$$
\vspace{-1.2cm}
\caption{Bifurcation diagrams for $a=3$ and $\kappa=0.9$.}
\label{figa3k.9asym}
$$\epsfxsize=9cm
\epsfysize=5.5cm
\epsfbox{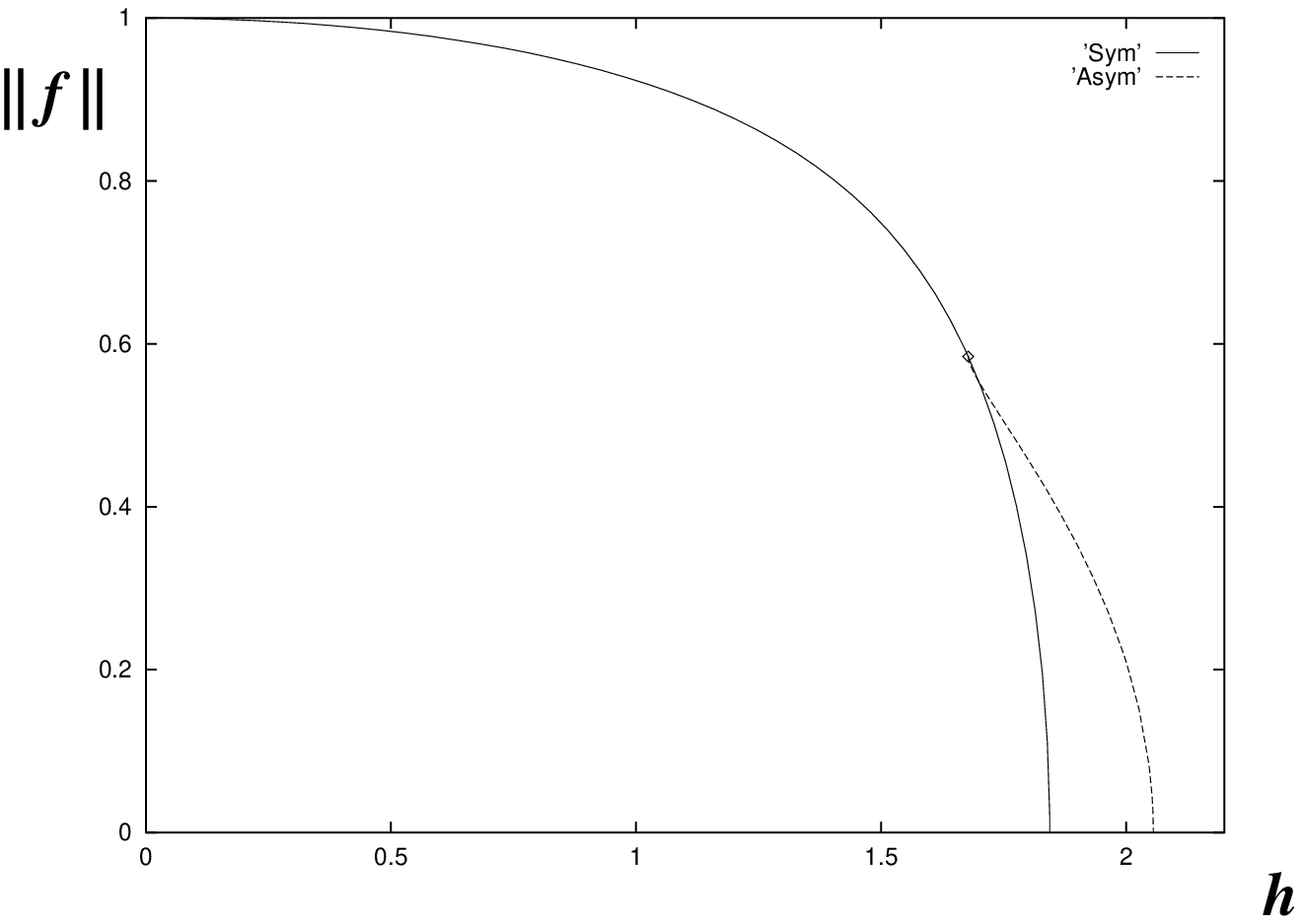}$$
\vspace{-1.2cm}
\caption{Bifurcation diagrams for $a=1$ and $\kappa=1.2$.}
\label{figa1k1.2}
\end{figure}

Recall that a fold in the symmetric branch correponds to a superheating phenomenon, a subcritical bifurcation at the highest nucleation field to a supercooling phenomenon and a supercritical bifurcation of asymmetric solution to surface superconductivity. In particular, when surface superconductivity occurs there is no
supercooling. 
 Thus, our physical interpretation of our new regions in the light of the discussion
 given by Tinkham \cite{T} is the following: 

\noindent
* $\kappa\leq \kappa_4(a)$ or $\kappa\geq \kappa_1(a)$ and $a\leq a^*$, that is $S_1 \cap A_0$ (Figure 1): thin films with type I behaviour. There is no supercooling nor superheating.

\noindent
* $\kappa> \kappa_4(a)$ and $\kappa> \kappa_1(a)$ and $\kappa> \kappa_3(a)$, that is $S_1 \cap A_2$ (Figure 10): thin film with type II behaviour. There is no supercooling nor superheating, but there is surface superconductivity
 due to the presence of a supercritical branch of asymmetric solutions. 

\noindent
* $\kappa \leq \kappa_5(a)$ and $\kappa \leq \kappa_1(a)$, that is $S_2 \cap A_1$ and $S_2 \cap A_0$ (Figures 2 and 6): type I behaviour. These superconductors supercool, that is the bifurcated branches from the normal solution are subcritical and superconductivity is nucleated in the bulk.

\noindent
* $\kappa_5(a)< \kappa \leq \kappa_2(a)$ and $a\geq a^*$, that is $S_2 \cap A_2$ (Figure 7): the material has type I characteristics because at $h_s$ the
 bifurcation of symmetric solutions 
 is subcritical. However, the bifurcation of asymmetric solutions
 is supercritical, and this results in the formation  of
 surface sheaths of
 superconductivity. Thus, in this regime, Tinkham \cite{T}  has given the
 name Type I 1/2 to the superconductor.

\noindent
* $\kappa_2(a)< \kappa \leq \kappa_3(a)$, that is $S_3 \cap A_2$ (Figure 9):
 type II behaviour,
   with superheating and surface superconductivity.

Thus, we claim that the curve $\kappa_5(a)$ is the limiting curve separating type
 I behaviours and type I 1/2.

Note that surface sheath effects can be inhibited if the superconducting
 surface is coated with a normal metal. In this situation, the boundary condition $f'(\pm a)=0$ needs to be changed and replaced by $f'(\pm a)=- f(\pm a)/b$
 as introduced by de Gennes \cite{DG}. This is why the type of superconductor is determined by the behaviour of the bifurcated branch at $h_s$ and not at $h_{as}$.

\hfill

To our knowledge, the global behaviour of the asymmetric bifurcation curves have never
 been mentionned in the literature. In light of the discussion given above the following open problems have
physical importance. 

\noindent
* prove how asymmetric solutions arise from bifurcation from symmetric solutions;

\noindent
* prove that as soon as the asymmetric curve arises from a subcritical bifurcation from
 the normal solution, the curve has no fold;

\noindent
* prove that as soon as the asymmetric curve arises from a supercritical bifurcation from
 the normal solution, the curve has a unique fold.

\subsection{The quintuple point}

As mentionned earlier, our experiments reveal the existence of a second
 key-point, in addition to the triple point defined in Section 3. To our knowledge,  the existence
 of this point has not previously been reported in the literature.
 For our discussion of the properties of this point, we refer the reader to
 Figure \ref{figsummary}.
 First, we let $a^*$ be the unique value such that $\kappa_1(a^*)=1/ \sqrt 2$.
 We have $a^*\simeq 1.23$.
 Then our investigation shows that $(a^*,1/ \sqrt 2)$ is on the curve
 $\kappa_4(a)$ and is the end point of $\kappa_5(a)$. At the point itself,
 we observe that $h(\beta)$ is decreasing to $h_s=h^*\simeq 1.45$
 and there are no asymmetric solutions. However, if we let $(a,\kappa)$ vary from this point, we expect to see 5 different behaviours of the bifurcation
 diagram. Along a ray leading into $S_1\cap A_0$, the bifurcation diagram is the
 same as shown in Figure \ref{figS1} and there is no asymmetric solution.
 If $(a,\kappa)$ lies on a ray leading into $S_1\cap A_2$, the bifurcation diagram is the
 same as shown in Figure \ref{figa1k1.2}: the asymmetric solutions bifurcate
 from $h^*$ and the asymmetric branch has a fold, though it is not easy to see 
 it on the picture. Next, let $(a,\kappa)$ lie on a ray leading into $S_2\cap A_0$; the bifurcation diagram is as in Figure \ref{figS2} so that the fold in the 
 symmetric branch results from a subcritical bifurcation. If the ray leads into
   $S_2\cap A_1$, the bifurcation diagram is as in Figure \ref{figa3k.35} so that there is a fold in the 
 symmetric branch resulting from a subcritical bifurcation at $h_s$ and there is also
 an asymmetric branch, with no fold, bifurcating from $h_{as}$. Both $h_s$ and $h_{as}$ are very close to $h^*$. Finally, if $(a,\kappa)$ lies in $S_2\cap A_2$,
 the diagram is as in Figure \ref{figa3}, this time the branch of asymmetric solutions has a fold.

 As we have seen in the discussion following Conjecture \ref{conjhas}, the 
 asymmetric solutions which appear play an important
 role in 2D vortex formation for intermediate values of $a$ larger than $a^*$.
 In fact, our investigation
    indicates that their existence is due to
 a complicated bifurcation phenomenon which occurs at the  quintuple point.
 Thus, a complete mathematical investigation of bifurcation 
 phenomena occuring at the quintuple point needs to be carried out.

\subsection{$\kappa_4(a)$}

Along the curve $\kappa_4(a)$, there are only symmetric solutions.
 As $(a,\kappa)$ passes from $A_0$ to $A_1$ or $A_2$
 across $\kappa_4(a)$, the asymmetric solutions bifurcate from $\beta=0$, $h=h_{as}$
 and $h_{as}$ tends to $h_s$ as $(a,\kappa)$ approaches a point on $\kappa_4(a)$.

The nonexistence of asymmetric solutions bifurcating from the normal
 solution for $\kappa a$ small was obtained by Bolley-Helffer \cite{BH8}.
\begin{theo}\label{BHunderk4}
 (Bolley-Helffer, \cite{BH8})
 There is a constant $C_0$ such that for $\kappa a\leq C_0$, there
 exists a unique nontrivial $C^\infty$ curve $(f(.,\eps),q(.,\eps),h(\eps))$ of solutions of $(GL)$ bifurcating from a normal solution $(0,h_0(x+e),h_0)$. These
 solutions bifurcate from the particular normal solutions $(0,h_0 x , h_0)$ where
 $h_0$ is the eigenvalue of a spectral problem, and therefore are 
 symmetric solutions.
\end{theo}

The existence of asymmetric solutions bifurcating from the normal
 solution for $\kappa a$ large was obtained by Bolley-Helffer \cite{BH8}.
\begin{theo}\label{BHabovek4}
 (Bolley-Helffer, \cite{BH8})
 There exists a constant $C_1$ such that for $\kappa a\geq C_1$, there
 exist exactly three nontrivial $C^\infty$ curves $(f(.,\eps),q(.,\eps),$ $h(\eps))$ of solutions of $(GL)$ bifurcating from normal solutions $(0,h_0(x+e),h_0)$,  where
 $h_0$ is an eigenvalue of a spectral problem. 
 One curve of
 solutions starts from the particular normal solutions $(0,h_0 x,$ $h_0)$, and is a curve of 
 symmetric solutions. The two others correspond to $e=\overline{e}(\kappa,a)$
 and $e=-\overline{e}(\kappa,a)$ and are asymmetric solutions.
\end{theo}
These two Theorems only give local behaviour of the bifurcation curve
 near the normal solutions, but this scaling  is consistent with the behaviour of $\kappa_4(a)$ 
  which is $C/a$. 

A global treatment of the region below $\kappa_4(a)$ for
 the full Ginzburg-Landau system
 remains an important open problem.

\subsection{$\kappa_5(a)$}
 As we have seen, $\kappa_5(a)$ is an extremely important curve.
 Our study indicates that as $(a,\kappa)$ passes from $A_1$ to $A_2$
 across $\kappa_5(a)$, there is a change of bifurcation for the asymmetric branch 
  from subcritical in $A_1$ to supercritical
  in $A_2$. This behaviour of the asymmetric branch is similar to what happens
 to the symmetric branch across $\kappa_1(a)$ and $\kappa_2(a)$.

The following additionnal information has been obtained in \cite{BH8}
 about the asymptotic behaviour of  $\kappa_5(a)$ when $a$ is large.
\begin{theo}\label{K5BH} (Bolley-Helffer, \cite{BH8})
There exists a constant $\Sigma_0$ approximately equal to 0.4 such that
 for any $\eta>0$ there exists $C_2$ such that for $\kappa a>C_2$,
 $|\kappa-\Sigma_0|>\eta$, $e=\overline e(\kappa,a)$ and $h_0$ the eigenvalue of a spectral problem, then the
 solution of the form (\ref{bifurcatedsol}) bifurcating from the normal solution
 $(0,h_0(x+e),h_0)$ satisfies $(\kappa-\Sigma_0)(h(\eps)-h_0)<0$. In particular, if $\kappa<\Sigma_0$, the bifurcation is subcritical, and if $\kappa>\Sigma_0$ the bifurcation is supercritical.
\end{theo}
Notice that $\Sigma_0$ is what we call $\kappa_{as}$. In \cite{BH8}, $\Sigma_0$
 is given in terms of two integrals. This value already appears in physics
 papers (see \cite{SJDG} for instance) and it is known to separate different
 behaviours of type I superconductors.

\subsection{$a$ large}
The only rigorous result concerning the behaviour 
 of the asymmetric branch is the following, which can be illustrated by Figure
 \ref{figa3k.9asym}.
\begin{theo}\label{theoHT}(Hastings and Troy \cite{HT2})
 Let $\kappa>1/ \sqrt {2.01}$ be fixed. Then for $a$ large enough, 
 there is a range of $h_0$ for which there is no symmetric solution
      and yet there is an asymmetric solution.
\end{theo}
This Theorem implies in particular that in the $(a,\kappa)$ range mentionned,
 we have $h_{as}>\overline{h}$ and $h_{as}>h_s$. This gives a partial resolution to Conjecture \ref{conjhas}. The physical importance of this result will
 be discussed in Section 5.

\hfill

\begin{figure}[htb]
$$\epsfxsize=9cm
\epsfysize=5.5cm
\epsfbox{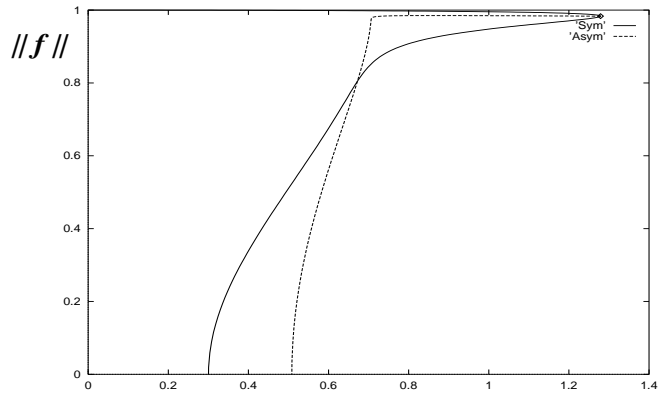}$$
\vspace{-1.2cm}
\caption{Bifurcation diagrams for $a=10$ and $\kappa=0.3$.}
\label{figa10k.3}
\end{figure}

Figure \ref{figa10k.3} illustrates the behaviour of the bifurcation diagram
 when $a$ is large. This is the limiting case of the one shown in Figure \ref{figa3k.35} when $a$ gets large. 
We see that the branching point of the asymmetric curve has the same limit as the fold of the symmetric curve. 
 Thus, for the limiting behaviour of the asymmetric curve when $a$ is large, our experiment
 lead us to the following.
\begin{conj}\label{conjasymalarge}
Let $\kappa$ be fixed. When $a$ tends to $\infty$,
 the limiting behaviour of the bifurcation curve of the asymmetric solutions
 is  the union of a horizontal line at $\beta=1$ from $h=1/\sqrt 2$ to $h=\overline{h}_\infty$
  (defined in Conjecture \ref{conj2S2infty}) and a curve joining $(h=1/ \sqrt 2,\beta=1)$
 to $(h=\kappa /\mu_1^0,\beta=0)$, where $\mu_1^0$ is approximately 0.59, and is defined as the minimum over $\alpha$ of the first eigenvalue of the
 Neumann problem of the harmonic oscillator in $(-\alpha,\infty)$. In particular,
\beq\label{limhb}
\lim_{a\to\infty}h_b= \overline{h}_\infty\quad \hbox{and}
\quad  \lim_{a\to\infty}\underline{h}_{as}=1/ \sqrt 2.
\eeq
\end{conj}

 This implies that $\kappa_{as}=\mu_1^0 / \sqrt 2$: when $\kappa<\kappa_{as}$, the curve from $\beta=0$ to $\beta=1$
 has positive slope resulting from a subcritical bifurcation
(see Figure \ref{figainfk.3}) and when $\kappa>\kappa_{as}$, this
 curve has negative slope resulting from a supercritical bifurcation (see Figures \ref{figainfk.5} and \ref{figainfk.9}). 
\begin{figure}[p]
\vspace{-.5cm}
\begin{center}
\input autoainfk.3.pstex_t
\end{center}
\vspace{-.9cm}
\caption{Limiting bifurcation diagrams for $a=\infty$ and $\kappa=0.3$.}
\label{figainfk.3}
\begin{center}
\input autoainfk.5.pstex_t
\end{center}
\vspace{-.9cm}
\caption{Limiting bifurcation diagrams for $a=\infty$ and $\kappa=0.5$.}
\label{figainfk.5}
\begin{center}
\input autoainfk.9.pstex_t
\end{center}
\vspace{-.9cm}
\caption{Limiting bifurcation diagrams for $a=\infty$ and $\kappa=0.9$.}
\label{figainfk.9}
\end{figure}

The value of $\mu_1^0 $ and the corresponding spectral problem were derived by Bolley-Helffer:
\begin{theo}\label{BHainftyasym} (Bolley-Helffer, \cite{BH1})
Let $\kappa$ be fixed. Then 
\beq\label{hasainfty}
\lim_{a\to\infty} h_{as}={\kappa \over {\mu_1^0}}.
\eeq
\end{theo}
This is consistent with the experimental value found by St-James and De Gennes \cite{SJDG} which is $2.392\kappa /\sqrt 2 $.

 

For fixed $\kappa$, Chapman, Howinson, McLeod and Ockendon \cite{CMLHO}
 prove the existence and uniqueness of the solution of 
 \beq\label{asym2}\left\{\begin{array}{ll}
{1 \over \kappa^2}f''=f(f^2+q^2-1)
\quad\hbox{in}\quad (-\infty,\infty),\\
\lim_{x\to-\infty}f(x)=1,\quad \lim_{x\to\infty}f(x)=0,\\\
q''=qf^2 \quad\hbox{in}\quad (-\infty,\infty),\\
\lim_{x\to\infty}q(x)=0,\quad \lim_{x\to\infty}q'(x)={1\over \sqrt 2},
\end{array}\right.\eeq
 This gives the point $\beta=1$, $h=1 / \sqrt 2$ in the bifurcation diagram, but further study is needed to resolve Conjecture \ref{conjasymalarge}.


\section{Stability of solutions}
\setcounter{equation}{0}

We define $(f,q)$ to be locally stable if it is a local minimizer
 of $E_\kappa$ in $H^1$ and unstable otherwise.
 Local and global stability properties of both symmetric and asymmetric solutions have received recent attention in the literature. As a result of our
 investigations, we make the following conjecture, which is to be read in the 
 light of Figure \ref{figsummary}.

\begin{conj}\label{conjstable}
In $S_1\cap A_0$, the symmetric solution is a global minimizer when it exists,
 that is for $0<h_0<h_s$ and the normal solution is a global minimizer for $h_0\geq h_s$.

In $S_2\cap A_0$ and in $S_2\cap A_1$,
  the symmetric solution is locally stable
 for $0<h_0<\overline{h}$, the normal solution is locally stable for $h_0\geq h_s$, and in $S_2\cap A_1$ the asymmetric solution is unstable. Moreover, there exists $h_c\in(h_s,\overline{h})$ such that
  the symmetric solution is a global minimizer for $0<h_0\leq h_c$
 and the normal solution is a global minimizer for $h_0\geq h_c$.

In $S_1\cap A_2$ the symmetric solution is locally stable
 for $0<h_0<h_b$, the asymmetric solution is locally stable
 for $\underline{h}_{as}<h_0<h_{as}$ and the normal solution is locally stable for $h_0\geq h_{as}$. Moreover, there exists $h_c\in(\underline{h}_{as},h_s)$ such that
  the symmetric solution is a global minimizer for $0<h_0\leq h_c$,
  the asymmetric solution is a global minimizer for $h_c\leq h_0 < h_{as}$
 and the normal solution is a global minimizer for $ h_0 \geq h_{as}$.

In $S_2\cap A_2$ the symmetric solution is locally stable
 for $0<h_0<\overline{h}$, the asymmetric solution is locally stable
 for $\underline{h}_{as}<h_0<h_{as}$ and the normal solution is locally stable for $h_0\geq h_{as}$. Moreover, there exists $h_c\in(h_s,\overline{h})$ such that
 the symmetric solution is a global minimizer for $0<h_0\leq h_c$,
 the asymmetric solution is a global minimizer for $h_c\leq h_0 < h_{as}$
 and the normal solution is a global minimizer for $h_0\geq h_{as}$.

In $S_3\cap A_2$ the symmetric solution is locally stable
 for $0<h_0<\overline{h}$ and locally unstable for $\underline{h}<h_0<h_s$, the asymmetric solution is locally stable
 for $\underline{h}_{as}<h_0<h_{as}$ and the normal solution is locally stable for $h_0\geq h_{as}$. Moreover, there exists $h_c\in(h_s,\overline{h})$ such that
  the upper symmetric solution is a global minimizer for $0<h_0\leq h_c$,
 the asymmetric solution is a global minimizer for $h_c\leq h_0 < h_{as}$
 and the normal solution is a global minimizer for $h_0\geq h_{as}$.
\end{conj}

This conjecture is consistent with physical observations of superheating
 and supercooling. The critical field $\overline{h}$ is often called superheating field because when starting from a symmetric superconducting solution near $h=0$, the material will remain in this locally stable state until $\overline{h}$ is reached. Then as $h_0$ is increased above $\overline{h}$, the maximum
 of $f$ has a discontinuity and bulk superconductivity is destroyed.

 Now, on the contrary, starting from  normal solution,
 when $h_0$ is decreased from infinity, the material remains in the normal state
 until the nucleation field ($h_s$ or $h_{as}$) is reached. If the branch of bifurcating solutions is unstable, this field is called supercooling field
 because nucleation occurs through a jump in the maximum of $f$: the material gets into the symmetric state which is the global
 minimizer.  These two phenomena give rise to hysteresis loops.

 For all the cases they have studied, Bolley-Helffer \cite{BH8} have established the local stability or instability of the branches bifurcating from normal
 solutions. In fact, when they proved the bifucation to be subcritical, they also proved local instability of the branch, and stability for supercritical branches, except in $S_3\cap A_2$ and $S_1\cap A_2$. Note that in $S_3\cap A_2$ and $S_1\cap A_2$, though the bifurcation is
 supercritical, they proved that the symmetric solution is unstable.

For $a$ fixed and $h_0\leq h_0^*(a)$ defined earlier in Section 3.5.2, 
 Aftalion \cite{aa1} proved that the global
 minimizer tends to a symmetric solution when $\kappa$ tends to
 infinity. Indeed, in this range corresponding to $S_1\cap A_2$, there are no asymmetric solutions for $h_0\leq h_0^*(a)$,
 but for $h_0> h_0^*(a)$, it is an interesting open problem to prove that
 both symmetric and asymmetric solutions coexist and the asymmetric
 solution is the global minimizer.

As we pointed out in Section 4, in $S_3\cap A_2$ and $S_2\cap A_2$, the formation of sheaths and vortices is related to the existence of stable
asymmetric solutions.  The Hastings-Troy result \cite{HT2}
gives a first step in the mathematical analysis of this phenomenon:
 Hastings and Troy \cite{HT2} gave a first step in the mathematical analysis of this phenomenon since they 
 proved that for fixed $\kappa>1/ \sqrt{2.01}$ and large $a$, there is a range of $h$ values for which
  the asymmetric solution is the global minimizer. As we have said before in Section 4, when the asymmetric solution is a global minimizer, it is not known for which regime of $(a,\kappa)$, it gives rise to surface sheath or to vortices.

\hfill

 We have provided a comprehensive study of the structure of
symmetric and asymmetric solutions for the entire range $0<a,\kappa<\infty$.
Throughout we have tied together our new observations with previously known
results. At the core of our new results are the existence of the triple point
and the quintuple point. These points provide the birth of families of
asymmetric and symmetric solutions via bifurcation. Already, one of the
authors, together with S.J. Chapman,
 has begun further investigations of the structure of both of
these  points and this has resulted in two new papers \cite{AC1}, \cite{AC2}.


\hfill

\noindent
{\bf Acknowledgement.} The authors are very grateful to P.J.McKenna for 
 mentionning the existence of AUTO.

\end{document}

%% file: autoka.pstex_t
\begin{picture}(0,0)%
\epsfig{file=autoka.pstex}%
\end{picture}%
\setlength{\unitlength}{0.00083300in}%
\begingroup\makeatletter\ifx\SetFigFont\undefined
\def\x#1#2#3#4#5#6#7\relax{\def\x{#1#2#3#4#5#6}}%
\expandafter\x\fmtname xxxxxx\relax \def\y{splain}%
\ifx\x\y   
\gdef\SetFigFont#1#2#3{%
  \ifnum #1<17\tiny\else \ifnum #1<20\small\else
  \ifnum #1<24\normalsize\else \ifnum #1<29\large\else
  \ifnum #1<34\Large\else \ifnum #1<41\LARGE\else
     \huge\fi\fi\fi\fi\fi\fi
  \csname #3\endcsname}%
\else
\gdef\SetFigFont#1#2#3{\begingroup
  \count@#1\relax \ifnum 25<\count@\count@25\fi
  \def\x{\endgroup\@setsize\SetFigFont{#2pt}}%
  \expandafter\x
    \csname \romannumeral\the\count@ pt\expandafter\endcsname
    \csname @\romannumeral\the\count@ pt\endcsname
  \csname #3\endcsname}%
\fi
\fi\endgroup
\begin{picture}(5824,4392)(289,-4141)
\put(376,-286){\makebox(0,0)[lb]{\smash{\SetFigFont{20}{24.0}{rm}$\kappa$}}}
\put(376,-1846){\makebox(0,0)[lb]{\smash{\SetFigFont{10}{24.0}{rm}$1/ \sqrt 2$}}}
\put(1950,-4011){\makebox(0,0)[lb]{\smash{\SetFigFont{10}{24.0}{rm}$\sqrt 5/ 2$}}}
\put(3451,-2911){\makebox(0,0)[lb]{\smash{\SetFigFont{20}{24.0}{bf}$S_2$}}}
\put(4801,-736){\makebox(0,0)[lb]{\smash{\SetFigFont{20}{24.0}{bf}$S_3$}}}
\put(6301,-4111){\makebox(0,0)[lb]{\smash{\SetFigFont{20}{24.0}{bf}$a$}}}
\put(5401,-1786){\makebox(0,0)[lb]{\smash{\SetFigFont{14}{24.0}{rm}$\kappa_2(a)$}}}
\put(2776,-661){\makebox(0,0)[lb]{\smash{\SetFigFont{14}{24.0}{rm}$\kappa_3(a)$}}}
\put(1576,-736){\makebox(0,0)[lb]{\smash{\SetFigFont{20}{24.0}{bf}$S_1$}}}
\put(1726,-2836){\makebox(0,0)[lb]{\smash{\SetFigFont{14}{24.0}{rm}$\kappa_1(a)$}}}
\end{picture}

%% file: autoAkappaa.pstex_t
\begin{picture}(0,0)%
\epsfig{file=autoAkappaa.pstex}%
\end{picture}%
\setlength{\unitlength}{0.00083300in}%
\begingroup\makeatletter\ifx\SetFigFont\undefined
\def\x#1#2#3#4#5#6#7\relax{\def\x{#1#2#3#4#5#6}}%
\expandafter\x\fmtname xxxxxx\relax \def\y{splain}%
\ifx\x\y   
\gdef\SetFigFont#1#2#3{%
  \ifnum #1<17\tiny\else \ifnum #1<20\small\else
  \ifnum #1<24\normalsize\else \ifnum #1<29\large\else
  \ifnum #1<34\Large\else \ifnum #1<41\LARGE\else
     \huge\fi\fi\fi\fi\fi\fi
  \csname #3\endcsname}%
\else
\gdef\SetFigFont#1#2#3{\begingroup
  \count@#1\relax \ifnum 25<\count@\count@25\fi
  \def\x{\endgroup\@setsize\SetFigFont{#2pt}}%
  \expandafter\x
    \csname \romannumeral\the\count@ pt\expandafter\endcsname
    \csname @\romannumeral\the\count@ pt\endcsname
  \csname #3\endcsname}%
\fi
\fi\endgroup
\begin{picture}(4164,2967)(289,-2716)
\put(402,-115){\makebox(0,0)[lb]{\smash{\SetFigFont{14}{16.8}{bf}$\kappa$}}}
\put(4350,-2696){\makebox(0,0)[lb]{\smash{\SetFigFont{14}{16.8}{bf}$a$}}}
\put(3338,-1937){\makebox(0,0)[lb]{\smash{\SetFigFont{14}{16.8}{bf}$A_1$}}}
\put(3444, 27){\makebox(0,0)[lb]{\smash{\SetFigFont{14}{16.8}{bf}{\small $\kappa_4(a)$}}}}
\put(1971,-419){\makebox(0,0)[lb]{\smash{\SetFigFont{14}{16.8}{bf}$A_2$}}}
\put(1111,-1532){\makebox(0,0)[lb]{\smash{\SetFigFont{14}{16.8}{bf}$A_0$}}}
\put(3444,-176){\makebox(0,0)[lb]{\smash{\SetFigFont{14}{16.8}{bf}{\small $\kappa_5(a)$}}}}
\end{picture}

%% file: autokafig.pstex_t
\begin{picture}(0,0)%
\epsfig{file=autokafig.pstex}%
\end{picture}%
\setlength{\unitlength}{0.00083300in}%
\begingroup\makeatletter\ifx\SetFigFont\undefined
\def\x#1#2#3#4#5#6#7\relax{\def\x{#1#2#3#4#5#6}}%
\expandafter\x\fmtname xxxxxx\relax \def\y{splain}%
\ifx\x\y   
\gdef\SetFigFont#1#2#3{%
  \ifnum #1<17\tiny\else \ifnum #1<20\small\else
  \ifnum #1<24\normalsize\else \ifnum #1<29\large\else
  \ifnum #1<34\Large\else \ifnum #1<41\LARGE\else
     \huge\fi\fi\fi\fi\fi\fi
  \csname #3\endcsname}%
\else
\gdef\SetFigFont#1#2#3{\begingroup
  \count@#1\relax \ifnum 25<\count@\count@25\fi
  \def\x{\endgroup\@setsize\SetFigFont{#2pt}}%
  \expandafter\x
    \csname \romannumeral\the\count@ pt\expandafter\endcsname
    \csname @\romannumeral\the\count@ pt\endcsname
  \csname #3\endcsname}%
\fi
\fi\endgroup
\begin{picture}(5824,4292)(289,-4141)
\put(451,-286){\makebox(0,0)[lb]{\smash{\SetFigFont{20}{24.0}{bf}$\kappa$}}}
\put(6301,-4111){\makebox(0,0)[lb]{\smash{\SetFigFont{20}{24.0}{bf}$a$}}}
\put(4801,-3661){\makebox(0,0)[lb]{\smash{\SetFigFont{14}{24.0}{bf}$S_2\cap A_0$}}}
\put(4801,-2986){\makebox(0,0)[lb]{\smash{\SetFigFont{14}{24.0}{bf}$S_2\cap A_1$}}}
\put(4801,-2461){\makebox(0,0)[lb]{\smash{\SetFigFont{14}{24.0}{bf}$S_2\cap A_2$}}}
\put(4801,-1561){\makebox(0,0)[lb]{\smash{\SetFigFont{14}{24.0}{bf}$S_3\cap A_2$}}}
\put(2026,-811){\makebox(0,0)[lb]{\smash{\SetFigFont{14}{24.0}{bf}$S_1\cap A_2$}}}
\put(1126,-2161){\makebox(0,0)[lb]{\smash{\SetFigFont{14}{24.0}{bf}$S_1\cap A_0$}}}
\put(5401, -16){\makebox(0,0)[lb]{\smash{\SetFigFont{5}{6.0}{bf}$\kappa_1(a)$}}}
\put(5401,-136){\makebox(0,0)[lb]{\smash{\SetFigFont{5}{6.0}{bf}$\kappa_2(a)$}}}
\put(5401,-266){\makebox(0,0)[lb]{\smash{\SetFigFont{5}{6.0}{bf}$\kappa_3(a)$}}}
\put(5401,-366){\makebox(0,0)[lb]{\smash{\SetFigFont{5}{6.0}{bf}$\kappa_4(a)$}}}
\put(5401,-486){\makebox(0,0)[lb]{\smash{\SetFigFont{5}{6.0}{bf}$\kappa_5(a)$}}}
\end{picture}

%% file: autoainfk.3.pstex_t
\begin{picture}(0,0)%
\epsfig{file=autoainfk.3.pstex}%
\end{picture}%
\setlength{\unitlength}{0.00083300in}%
\begingroup\makeatletter\ifx\SetFigFont\undefined
\def\x#1#2#3#4#5#6#7\relax{\def\x{#1#2#3#4#5#6}}%
\expandafter\x\fmtname xxxxxx\relax \def\y{splain}%
\ifx\x\y   
\gdef\SetFigFont#1#2#3{%
  \ifnum #1<17\tiny\else \ifnum #1<20\small\else
  \ifnum #1<24\normalsize\else \ifnum #1<29\large\else
  \ifnum #1<34\Large\else \ifnum #1<41\LARGE\else
     \huge\fi\fi\fi\fi\fi\fi
  \csname #3\endcsname}%
\else
\gdef\SetFigFont#1#2#3{\begingroup
  \count@#1\relax \ifnum 25<\count@\count@25\fi
  \def\x{\endgroup\@setsize\SetFigFont{#2pt}}%
  \expandafter\x
    \csname \romannumeral\the\count@ pt\expandafter\endcsname
    \csname @\romannumeral\the\count@ pt\endcsname
  \csname #3\endcsname}%
\fi
\fi\endgroup
\begin{picture}(3708,2689)(309,-2441)
\put(410,-79){\makebox(0,0)[lb]{\smash{\SetFigFont{12}{14.4}{rm}$\beta$}}}
\put(3916,-2416){\makebox(0,0)[lb]{\smash{\SetFigFont{12}{14.4}{bf}$h$}}}
\put(1400,-2326){\makebox(0,0)[lb]{\smash{\SetFigFont{12}{14.4}{rm}$\kappa$}}}
\put(1759,-2326){\makebox(0,0)[lb]{\smash{\SetFigFont{12}{14.4}{rm}{\tiny 1.695}$\kappa$}}}
\put(2387,-2326){\makebox(0,0)[lb]{\smash{\SetFigFont{12}{14.4}{rm}{\tiny $1/\sqrt 2$}}}}
\end{picture}

%% file: autoainfk.5.pstex_t
\begin{picture}(0,0)%
\epsfig{file=autoainfk.5.pstex}%
\end{picture}%
\setlength{\unitlength}{0.00083300in}%
\begingroup\makeatletter\ifx\SetFigFont\undefined
\def\x#1#2#3#4#5#6#7\relax{\def\x{#1#2#3#4#5#6}}%
\expandafter\x\fmtname xxxxxx\relax \def\y{splain}%
\ifx\x\y   
\gdef\SetFigFont#1#2#3{%
  \ifnum #1<17\tiny\else \ifnum #1<20\small\else
  \ifnum #1<24\normalsize\else \ifnum #1<29\large\else
  \ifnum #1<34\Large\else \ifnum #1<41\LARGE\else
     \huge\fi\fi\fi\fi\fi\fi
  \csname #3\endcsname}%
\else
\gdef\SetFigFont#1#2#3{\begingroup
  \count@#1\relax \ifnum 25<\count@\count@25\fi
  \def\x{\endgroup\@setsize\SetFigFont{#2pt}}%
  \expandafter\x
    \csname \romannumeral\the\count@ pt\expandafter\endcsname
    \csname @\romannumeral\the\count@ pt\endcsname
  \csname #3\endcsname}%
\fi
\fi\endgroup
\begin{picture}(3627,2692)(289,-2441)
\put(410,-79){\makebox(0,0)[lb]{\smash{\SetFigFont{12}{14.4}{rm}$\beta$}}}
\put(3916,-2416){\makebox(0,0)[lb]{\smash{\SetFigFont{12}{14.4}{bf}$h$}}}
\put(2287,-2311){\makebox(0,0)[lb]{\smash{\SetFigFont{12}{14.4}{rm}{\tiny $1/\sqrt 2$}}}}
\put(1876,-2311){\makebox(0,0)[lb]{\smash{\SetFigFont{12}{14.4}{rm}$\kappa$}}}
\put(2751,-2311){\makebox(0,0)[lb]{\smash{\SetFigFont{12}{14.4}{rm}{\tiny1.695}$\kappa$}}}
\end{picture}

%% file: autoainfk.9.pstex_t
\begin{picture}(0,0)%
\epsfig{file=autoainfk.9.pstex}%
\end{picture}%
\setlength{\unitlength}{0.00083300in}%
\begingroup\makeatletter\ifx\SetFigFont\undefined
\def\x#1#2#3#4#5#6#7\relax{\def\x{#1#2#3#4#5#6}}%
\expandafter\x\fmtname xxxxxx\relax \def\y{splain}%
\ifx\x\y   
\gdef\SetFigFont#1#2#3{%
  \ifnum #1<17\tiny\else \ifnum #1<20\small\else
  \ifnum #1<24\normalsize\else \ifnum #1<29\large\else
  \ifnum #1<34\Large\else \ifnum #1<41\LARGE\else
     \huge\fi\fi\fi\fi\fi\fi
  \csname #3\endcsname}%
\else
\gdef\SetFigFont#1#2#3{\begingroup
  \count@#1\relax \ifnum 25<\count@\count@25\fi
  \def\x{\endgroup\@setsize\SetFigFont{#2pt}}%
  \expandafter\x
    \csname \romannumeral\the\count@ pt\expandafter\endcsname
    \csname @\romannumeral\the\count@ pt\endcsname
  \csname #3\endcsname}%
\fi
\fi\endgroup
\begin{picture}(3627,2694)(289,-2443)
\put(410,-79){\makebox(0,0)[lb]{\smash{\SetFigFont{12}{14.4}{rm}$\beta$}}}
\put(3916,-2416){\makebox(0,0)[lb]{\smash{\SetFigFont{12}{14.4}{bf}$h$}}}
\put(1876,-2386){\makebox(0,0)[lb]{\smash{\SetFigFont{12}{14.4}{rm}{\tiny $1/\sqrt 2$}}}}
\put(2326,-2386){\makebox(0,0)[lb]{\smash{\SetFigFont{12}{14.4}{rm}$\kappa$}}}
\put(3301,-2386){\makebox(0,0)[lb]{\smash{\SetFigFont{12}{14.4}{rm}{\tiny 1.695}$\kappa$}}}
\end{picture}